\documentclass[journal]{IEEEtran}

\usepackage{amsmath}
\usepackage{graphicx}
\usepackage{subfig} 
\usepackage{tabu}  

\usepackage[colorlinks,linkcolor=blue,anchorcolor=blue,citecolor=green]{hyperref}
\usepackage{multirow}
\usepackage{dsfont}
\usepackage{stfloats}
\usepackage{amsmath}
\usepackage{cite}
\usepackage{booktabs}
\usepackage{algorithm}
\usepackage{algorithmic}
\usepackage{amssymb}
\ifCLASSINFOpdf

\else
 
\fi

\begin{document}

\title{Unsupervised Change Detection in Multi-temporal VHR Images Based on Deep Kernel PCA Convolutional Mapping Network}

\author{
        Chen~Wu,~\IEEEmembership{Member,~IEEE,}
        Hongruixuan~Chen,
        Bo~Du,~\IEEEmembership{Senior~Member,~IEEE,}
        and~Liangpei~Zhang,~\IEEEmembership{Fellow,~IEEE}

\thanks{Manuscript submitted December 14, 2019. This work was supported in part by the National Natural Science Foundation of China under Grant 61971317, 41801285, 61822113 and 41871243.}
\thanks{C. Wu is with the State Key Laboratory of Information Engineering in Surveying, Mapping and Remote Sensing, and School of Computer Science, Wuhan University, Wuhan, P.R. China (e-mail: chen.wu@whu.edu.cn).}
\thanks{H. Chen is with the State Key Laboratory of Information Engineering in Surveying, Mapping and Remote Sensing, Wuhan University, Wuhan, P.R. China (e-mail:Qschrx@whu.edu.cn).}
\thanks{B. Du is with the School of Computer Science, and Collaborative Innovation Center of Geospatial Technology, Wuhan University, Wuhan, P.R. China (email: gunspace@163.com).}
\thanks{L. Zhang is with the Remote Sensing Group, State Key Laboratory of Information Engineering in Surveying, Mapping, and Remote Sensing, Wuhan University, Wuhan, P.R. China (e-mail: zlp62@whu.edu.cn, corresponding author).}
}

\markboth{SUBMITTED TO IEEE TRANSACTIONS ON CYBERNETICS ON December 14, 2019}%
{Shell \MakeLowercase{\textit{et al.}}: Bare Demo of IEEEtran.cls for IEEE Journals}

\maketitle

\begin{abstract}
  With the development of Earth observation technology, very-high-resolution (VHR) image has become an important data source of change detection. Nowadays, deep learning methods have achieved conspicuous performance in the change detection of VHR images. Nonetheless, most of the existing change detection models based on deep learning require annotated training samples. In this paper, a novel unsupervised model called kernel principal component analysis (KPCA) convolution is proposed for extracting representative features from multi-temporal VHR images. Based on the KPCA convolution, an unsupervised deep siamese KPCA convolutional mapping network (KPCA-MNet) is designed for binary and multi-class change detection. In the KPCA-MNet, the high-level spatial-spectral feature maps are extracted by a deep siamese network consisting of weight-shared PCA convolution layers. Then, the change information in the feature difference map is mapped into a 2-D polar domain. Finally, the change detection results are generated by threshold segmentation and clustering algorithms. All procedures of KPCA-MNet does not require labeled data. The theoretical analysis and experimental results demonstrate the validity, robustness, and potential of the proposed method in two binary change detection data sets and one multi-class change detection data set. 
\end{abstract}


\IEEEpeerreviewmaketitle

\section{Introduction}\label{sec:1}

\IEEEPARstart{T}{he} Earth surface is constantly evolving. Real-time and accurate access to the surface changes is of great importance for the better understanding of human activities, ecosystem and their interactions \cite{Lu2004}. Multi-temporal remote sensing images can reveal the dynamic changes in the surface. Therefore, change detection based on multi-temporal images has become more and more significant. Change detection (CD) is the process of identifying differences in the state of an object or phenomenon by observing it at different time \cite{Singh1989}, which has played an important role in a large number of applications, to name a few, land-use and land-cover change analysis, resource management, ecosystem monitoring, and disaster assessment \cite{Xian2009,Coppin2004,Luo2018,Sim2006,Zelinski2014,2014Li,Brunner2010,Wu2017b,Doroodgar2014}. Recently, owing to the development of Earth observation technology, very-high-resolution (VHR) images are available by more and more satellite sensors (e.g. SPOT, IKONOS, QuickBird, and GaoFen). The VHR images can provide abundant surface details and spatial distribution information, which makes it possible to detect subtler changes. Now, CD in multi-temporal VHR images has caught more and more attention \cite{Bovolo2009,Somasundaram2010,Chen2012,Lei2014,Huo2016,Tan2016,Zhan2017,Lv2018a,Daudt2018,Security2018,Wu2017a}. 

\par Pixel-based change detection (PBCD) model is the earliest CD method, which has been deeply studied and widely used \cite{Singh1989,Sharma2007,Bovolo2012,Thonfeld2016,Deng2008, Nielsen1997,Nielsen2007,Wu2014}. Change vector analysis (CVA) \cite{Sharma2007} is a classic CD model, which is designed to calculate change intensity and change direction for binary and multi-class CD. Based on CVA, many extension models are proposed, such as compressed CVA ($C^{2}$VA) \cite{Bovolo2012}, robust CVA (RCVA) \cite{Thonfeld2016} and so on. Principal component analysis (PCA) is a transformation method, which transforms the difference images or stacked images into a new feature space and selects a part of principal components for CD \cite{Deng2008}. In \cite{Nielsen1997} and \cite{Nielsen2007}, Nilsen et al propose multivariate alteration detection (MAD) that aims to maximize the variance of transformed variables and is invariant to affine transformation. Based on the theory of slow feature analysis (SFA), Wu et al \cite{Wu2014} propose a novel CD model that is able to extract the most invariant components from multi-temporal images, and transform original images into a new feature space, where the changed pixels are highlighted and unchanged ones are suppressed. Though these PBCD models are effective in their respective application scenarios, they only utilize spectral information of multi-temporal images due to the limitation of models themselves. However, the increase in the spatial resolution of VHR images limits the spectral resolution and brings the spectral variability, which makes CD results of PBCD models suffered from the salt-and-pepper noise and internal fragmentation \cite{Huo2016,Lv2018a,Li2017b}. Therefore, it is greatly important to employ the spatial context information and spectral information in VHR images CD. 

\par For the purpose of exploring the spatial context information in VHR images, two major and conventional approaches, object-based change detection (OBCD) model and spatial feature extraction method are developed. In OBCD, the image object consisting of pixels with similar spectral signatures becomes the basic unit of CD \cite{Hussain2013,Gil-Yepes2016,Desclee2006,Somasundaram2010}. To get the image objects, the first step of OBCD is image segmentation, which divides the images into multiple homogeneous regions. After obtaining the image objects, some representative features of objects, such as shape index, texture index, and mean and standard deviation of each band, are extracted for change detection. The second approach is the spatial feature extraction methods. In these methods, spatial context information is extracted from the local area of each pixel In \cite{Lu2017,Lei2014,Hoberg2015,Zhou2016}. In \cite{Lei2014}, a CD method based on texton forest is proposed to capture spatial context information in VHR images. Integrating macro- and micro-texture features with random forest and fuzzy set model, Li et al \cite{Li2017c} propose a multi-texture CD method to extract spatial features for CD. Hoberg el at \cite{Hoberg2015} introduce conditional random field (CRF) model into CD of multi-temporal VRH images to model spatial information in VHR images and much research is conducted after that \cite{Zhou2016,Lv2016,Cao2016,Lv2018a}. Nonetheless, in both aforementioned approaches, only low-level features are utilized and many of them are hand-crafted, which are insufficient for representing the key information of original data and coping with complex ground situations in VHR images. In addition, the performance of OBCD depends on the effects of segmentation, but sometimes it is difficult to select the appropriate segmentation parameters. 

\par Last few years, deep learning (DL) has achieved spectacular performance in the field of computer vision and remote sensing image interpretation \cite{Lecun2015,Zhang2016b,Zhu2017a}. Different from conventional methods, DL models have the capability to extract representative high-level features from VHR images. Therefore, a variety of models based DL are proposed for CD in multi-temporal VHR images. In \cite{Gong2016}, a CD method based on deep neural network (DNN) is proposed for CD in SAR images. In \cite{Zhan2017}, Zhan et al specifically design a deep siamese convolutional network for multi-temporal aerial images, which extracts spatial-spectral features by two weight-shared branches. Lyu et al \cite{Lyu2016} adopt recurrent neural network (RNN) to tackle the temporal connection between multi-temporal images. Going one step further, in \cite{Mou2019}, a CD architecture based on recurrent convolutional neural network is proposed to extract unified features for binary and multi-class CD. Combining a pre-trained deep convolutional neural network with CVA, Saha et al \cite{Saha2019} design a CD method called deep CVA for VHR images CD. In \cite{CayeDaudt2018}, Daudt et al first introduce fully convolutional network (FCN) into CD and propose two siamese extensions of FCN, which achieve good performance in two open VHR images CD datasets. Though these methods based on DL models achieve good performance in CD, the training process of DL models is in supervised learning fashion with annotated data. And it is undeniable that the manual selection of annotated samples is labor-consuming, especially for remote sensing data. 

\par Therefore, several unsupervised feature extraction models, including restricted Boltzmann machines (RBMs) \cite{hinton2006} and auto-encoder (AE) \cite{Bengio2007}, have been adopted to solve this problem. However, these models flatten the image patches into vectors, ignoring the property of image in the spatial domain, due to equipped with fully connected layers \cite{Mou2019}. Another way is cooperating deep learning models with unsupervised pre-detection algorithms. Gao et al \cite{Gao2016} present a CD method based on PCANet \cite{Chan2015} for multi-temporal SAR images, and a pre-detection algorithm based on Gabor wavelets and fuzzy c-means is utilized to select interested pixels with a high probability of being changed or unchanged. Then the network is trained on the samples selected by the automatic pre-detection algorithm. Learning nonlinear features with DNN and highlighting changes via SFA, Ru et al \cite{Du2019a} propose an unsupervised deep slow feature analysis (DSFA) model for CD. For training the DNN of DSFA, a pre-detection method based on CVA is utilized to selecting samples. In \cite{Li2019a}, a supervised spatial fuzzy clustering is adopted to produce pseudo-labels for training the DCNN. This approach solves the sample problem of the DL model to a certain degree. However, if the pre-detection algorithm does not perform well on one data set, the performance of DL model is also damaged. What’s more, most of these existing DL-based methods are merely focus on binary CD. And there are currently only a few methods \cite{Saha2019,Zhang2019} that can be used for unsupervised multi-class CD. 

\par Considering the above issues comprehensively, in this paper, we utilize the unsupervised subspace learning algorithms, kernel principal component analysis (KPCA) \cite{Bernhard1998}, to develop a novel feature extraction model called KPCA convolution to extract representative spatial-spectral features from VHR images in a totally unsupervised manner. Based on the KPCA convolution, a powerful and general network called KPCA-MNet is designed for unsupervised binary and multi-class CD. First, the high-level spatial-spectral feature maps are extracted by KPCA-MNet with deep siamese network architecture. Then, pixel-wise subtraction is implemented to get the feature difference map. To efficiently utilizing the change information in the feature difference map, KPCA-MNet maps the feature difference map into a 2-D polar domain. Finally, the unsupervised threshold segmentation methods or clustering techniques would be performed to get the desired CD results. 

\par The rest of this paper is organized as follows. In section \ref{sec:2}, the background of KPCA and CNN are introduced. Section \ref{sec:3} elaborates the proposed KPCA convolution and KPCA-MNet. Section \ref{sec:4} provides experimental settings, experimental results and discussion. In section \ref{sec:5}, the experiment of multi-class CD is carried. Finally, Section \ref{sec:6} draws the conclusion of our work in this paper. 

\section{Background}\label{sec:2}
\subsection{KPCA}
\par PCA is a powerful technique for feature extraction that uses an orthogonal transformation to convert original data into new feature space \cite{Wold1987}. In the new feature space, possibly correlated data becomes linearly uncorrelated. 

\par But in some case, the principal components in input space which are linear related to the input variables cannot extract representative features from the observation data well. The PCA can be generalized to the KPCA, which maps the original data into a nonlinear high-dimensional space, to better extract the essential high-level representation of original observation data. 

\par Considering a possibly nonlinear map $\Phi:R^{N}\rightarrow H$, where $H$ is the feature space that could have an arbitrarily large, even infinite, dimensionality. The original data $x\rightarrow\Phi\left(x\right)$ could be mapped into a new nonlinear feature space $x$. The covariance matrix of mapped data is expressed as:
\begin{equation}
  C^{\Phi}=\frac{1}{M}\sum_{i=1}^{M}\tilde{\Phi}\left(x_{i}\right)\tilde{\Phi}^{T}\left(x_{i}\right),
\label{eq:1}
\end{equation} 
where $\tilde{\Phi}\left(x_{i}\right)$ is the centralized mapped data. The basic idea of KPCA is to solve the problem of PCA in the new nonlinear feature space $F$. This can be achieved by solving the following eigenvalue problem: 
\begin{equation}
  \lambda^{\Phi}v^{\Phi}=C^{\Phi}v^{\Phi}.
\label{eq:2}
\end{equation} 
\par Since the eigenvectors can be expressed as the linear combination of mapped data \cite{Bernhard1998}
\begin{equation}
  v^{\Phi}=\sum_{i=1}^{M}\alpha_{i}\tilde{\Phi}\left(x_{i}\right),
\label{eq:3}
\end{equation} 
and considering
\begin{equation}
  \lambda^{\Phi}\left(v^{\Phi}\right)^{T}\tilde{\Phi}\left(x_{j}\right)=\left(v^{\Phi}\right)^{T}C^{\Phi}\tilde{\Phi}\left(x_{j}\right),
\label{eq:4}
\end{equation} 
the Eq. (2) can be reformulated as
\begin{equation}  
  \begin{split}
    \lambda^{\Phi}&\sum_{i=1}^{M}\alpha_{i}\tilde{\Phi}^{T}\left(x_{i}\right)\tilde{\Phi}\left(x_{j}\right)= \\
         &\frac{1}{M}\sum_{i=1}^{M}\alpha_{i}\tilde{\Phi}^{T}\left(x_{i}\right)\left(\sum_{k=1}^{M}\tilde{\Phi}\left(x_{j}\right)\tilde{\Phi}^{T}\left(x_{k}\right)\right)\tilde{\Phi}\left(x_{j}\right)
  \end{split}
\label{eq:5}
\end{equation}

\par Defining define an $M \times M$ kernel matrix $\tilde{K}$ as
\begin{equation}
  \tilde{K}_{ij}=\left(\tilde{\Phi}\left(x_{i}\right)\cdot\tilde{\Phi}\left(x_{j}\right)\right).  
\label{eq:6}
\end{equation} 
Therefore, the eigenvalue problem of KPCA can be rewritten as
\begin{equation}
  M\lambda^{\Phi}\tilde{K}\alpha=\tilde{K}^{2}\alpha.  
\label{eq:7}
\end{equation} 
As $K$ is a symmetric matrix, it has a set of eigenvectors which spans the whole space \cite{Bernhard1998}, thus
\begin{equation}
  M\lambda^{\Phi}\alpha=\tilde{K}\alpha.  
\label{eq:8}
\end{equation} 
can give all solutions $\alpha$ of Eq. (7). Further, we can normalize $\alpha$ to make the corresponding eigenvectors in $H$ be normalized
\begin{equation}
  \left(v^{\Phi}_{k}\cdot v^{\Phi}_{k}\right)=\lambda\left(\alpha_{k}\cdot \alpha_{k}\right),k=1,2,\cdots,M.
\label{eq:9}
\end{equation} 

\par For the purpose of feature extraction, it should compute projections of original data on the eigenvectors $v^{\Phi}$ in $H$. Let $y$ be a test data, with an image $\Phi\left(y\right)$ in $H$, the $k$-th feature component can be calculated as
\begin{equation}
  \left(v^{\Phi}_{k}\right)^{T}\Phi\left(y\right)=\sum_{i=1}^{M}\alpha^{i}_{k}\left(\tilde{\Phi}\left(x_{i}\right)\cdot \Phi\left(y\right)\right),k=1,2,\cdots,M.
\label{eq:10}
\end{equation} 

\par In the above procedure, although we assume that $\tilde{\Phi}\left(x\right)$ is centralized, it is actually difficult to compute the mean of the mapped data directly in $H$. Therefore, in KPCA, the centralized kernel matrix is computed by the original kernel matrix \cite{Bernhard1998}
\begin{equation}
  \tilde{K}_{ij}=\left(K-\frac{1}{M}1K-\frac{1}{M}K1+\frac{1}{M^2}1K1\right)_{ij},
\label{eq:11}
\end{equation} 
where \textbf{1} is an $m\times m$ matrix with elements of \textbf{1}, $K$ is the original kernel matrix directly calculated by original nonlinear mapped data  $\Phi\left(x\right)$.

\par Note that in the procedure of KPCA, we only need to compute the dot products of the nonlinear mapped data, namely  $\left(\Phi\left(x\right)\cdot \Phi\left(y\right)\right)$. Therefore, the kernel function can be used to compute their dot products in $F$ without explicitly calculating the nonlinear mapped data $\Phi\left(x\right)$ \cite{boser1992}. There are several commonly used kernel functions in KPCA, including polynomial kernel, radial basis function (RBF), sigmoid kernel, and cosine kernel. 

\subsection{CNN}
\par Convolutional neural network is a specialized kind of neural network for processing data that has a known, grid-like topology \cite{Lecun2015}, such as image data. Now, CNN has been tremendously successful in a variety of computer vision tasks \cite{Lecun2015,He2016,Han2018,Chen2018,Li2019c} and become one of the most popular network architectures. 

\par Generally, CNN is composed of several convolutional layers, followed by activation functions, and pooling layers. Given an image or other grid-like topology data, a set of feature maps are extracted by multiple trainable convolution kernel in convolutional layer, i.e. $F=C\left(I\right)$
\begin{equation}
  F^{\left(i\right)}=C^{\left(i\right)}\left(I\right)=a\left(W^{\left(i\right)}*I+b^{\left(i\right)} \right)
\label{eq:12}
\end{equation}
where $F^{\left(i\right)}$ is the $i$-th feature map obtained by the $i$-th convolution kernel $W^{\left(i\right)}$ and bias $b^{\left(i\right)}$, $*$ denotes convolution operation, and $a$ is the activation function following the convolution layer, which can introduce non-linearity into network. Then the pooling layer subsamples each feature map to remove redundant features, and helps to make the features approximately invariant to spatial translations. After processed by several alternating convolutional layers and pooling layers, the high-level features of input data are finally generated in an automatic manner instead of hand-crafted. 

\par The training (or optimization) process of CNN is in a supervised learning fashion with annotated data. In the training process, stochastic gradient descent with the backpropagation (BP) algorithm \cite{Y1998} is usually utilized to optimize the parameters in CNN. As the training process goes on, the features generated by the network are more and more representative.

\section{Methodology}\label{sec:3}
\subsection{KPCA Convolution for VHR Images}
\par Compared with low- and medium-resolution images, VHR images could not only provide spectral information, but also afford abundant ground details, texture features, and spatial distribution information. Therefore, the spatial-spectral feature is of great importance for change detection in VHR images. The convolutional layer of CNN is a suitable structure for simultaneously extracting spatial-context features and spectral features in a 2-D manner. However, as we mentioned in section II-B, the training process of CNN is in a supervised learning fashion with annotated data. And in the field of remote sensing, datasets with ground-truth labels are not always available in practical applications. 

\par In this subsection, a novel convolution operation based on KPCA is proposed to extract the representative spatial-spectral features from VHR images in an unsupervised manner. Let us consider the case of linear PCA first. 

\subsubsection{PCA Convolution}
Given a VHR image $I\in R^{h\times w\times c}$, around each pixel, a corresponding $s_{1}\times s_{2}\times c$ image patches is taken. For the entire image, all image patches are vectorized and collected, i.e. $p_{1},p_{2},\cdots,p_{hw}\in R^{s_{1}s_{2}c}$. Then, $n$ vectorized image patches are randomly selected as training samples, we denote $P_{train}=\left[p^{train}_{1},p^{train}_{2},\cdots,p^{train}_{n}\right]\in R^{s_{1}s_{2}c\times n}$.

\par The covariance matrix of the training samples can be calculated as

\begin{equation}
  \begin{split}
  C&=\frac{1}{n}P_{train}P^{T}_{train} \\
  &=\frac{1}{n}\sum^{n}_{i=1}p_{i}^{train}\left(p_{i}^{train}\right)^{T}\in R^{s_{1}s_{2}c\times s_{1}s_{2}c}.
\end{split}
\label{eq:13} 
\end{equation}
The eigenvalues $\lambda_{1}\geq\lambda_{2}\geq\cdots\geq\lambda_{s_{1}s_{2}c}$ and corresponding eigenvectors $v_{1},v_{2},\cdots,v_{s_{1}s_{2}c}\in R^{s_{1}s_{2}c}$ are available by solving the PCA problem. Utilizing the first $k$ components to transform image patches into a new feature space, the representative spatial-spectral features can be extracted as
\begin{equation}
  Y=U^{T}P,
\label{eq:14}
\end{equation}
where $U=\left[v_{1},v_{2},\cdots,v_{k}\right]$, $Y=\left[y_{1},y_{2},\cdots,y_{hw}\right]\in R^{k\times hw}$, and $y_{i}\in R^{k}$ is the spatial-spectral features of the $k$-th pixel. 

\par Note that the above procedure is equivalent to a convolution operation
\begin{equation}
  F_{j}=W_{j}*I, j=1,2,\cdots,k,
\label{eq:15}
\end{equation}
where the $i$-th convolution kernel $W_{j}\in R^{s_{1}\times s_{2}\times c}$ is the reshaped eigenvector $v_{j}$, and $F\in R^{h\times w}$ is the $j$-th channel of the spatial-spectral feature map $F$. 

\subsubsection{Kernel PCA Convolution}
\par As seen from Eq. (15), PCA convolution is a linear transformation that limits the representing ability of extracted features. To improve the representativeness and discriminability of the extracted features and subsequent CD performance, the non-linearity is introduced into spatial-spectral features by means of mapping original images into a new nonlinear feature space. To do this, PCA convolution should be generalized to KPCA convolution. 

\par Considering a nonlinear map, the vectorized image patches taken from the VHR image I can be mapped into a high-dimensional nonlinear feature space
\begin{equation}
  P\rightarrow \Phi\left(P\right)=\left[\Phi\left(p_{1}\right),\Phi\left(p_{2}\right),\cdots,\Phi\left(p_{n}\right)\right].
\label{eq:16}
\end{equation}

\par The covariance matrix of the training samples is expressed as
\begin{equation}
  C^{\Phi}=\tilde{\Phi}\left(P_{\text {train}}\right) \tilde{\Phi}^{T}\left(P_{\text {train}}\right)=\frac{1}{n} \sum_{i=1}^{n} \tilde{\Phi}\left(p_{i}^{\text {train}}\right) \tilde{\Phi}^{T}\left(p_{i}^{\text {train}}\right).
\label{eq:17}
\end{equation}

\par The nonlinear eigenvectors (principal components) can be achieved by solving the KPCA problem
\begin{equation}
  \lambda^{\Phi}v^{\Phi}=C^{\Phi}v^{\Phi}.
\label{eq:18}
\end{equation}

\par As elaborated in section II-A, the eigenvalue problem of KPCA can be rewritten as
\begin{equation}
  n \lambda^{\Phi} \alpha=\tilde{K} \alpha.
\label{eq:19}
\end{equation}
By solving this equation and get all solutions $\alpha$, according to Eq. (\ref{eq:9}) and Eq. (\ref{eq:10}), the nonlinear spatial-spectral features can be extracted by the first $k$ nonlinear principal components as
\begin{equation}
  \begin{split}
  F_{j}^{\Phi}&=\left(v_{j}^{\Phi}\right)^{T} \Phi(P) \\
  &=\sum_{i=1}^{n} \alpha_{i}^{j}\left(\tilde{\Phi}\left(p_{i}^{t r a i n}\right) \cdot \Phi(P)\right), j=1,2, \cdots, k,
  \end{split}
  \label{eq:20}
\end{equation}
where $F_{j}\in R^{hw}$ is the $j$-th channel of vectorized nonlinear spatial-spectral feature map. As seen from Eq. (\ref{eq:20}), in the nonlinear feature extraction process, only the dot products of the nonlinear mapped data should be calculated. Therefore, a suitable kernel function $k^{\Phi}\left(x,y\right)$ is utilized to compute the dot products and extract nonlinear spatial-spectral features without directly employing the nonlinear mapping $\Phi$:
\begin{equation}
  F_{j}^{\Phi}=\left(v_{j}^{\Phi}\right)^{T} \Phi(P)=\sum_{i=1}^{n} \alpha_{i}^{j}k^{\Phi}\left(p^{train}_{i},P\right), j=1,2, \cdots, k.
\label{eq:21}
\end{equation}

\par The above procedure is also equivalent to a convolution operation, though the convolution kernels are not constructed explicitly
\begin{equation}
  F_{j}^{\Phi}=\varphi\left(W_{j}*I\right), j=1,2, \cdots, k
\label{eq:22}
\end{equation}
where $F_{i}^{\Phi}\in R^{h\times w}$ is the $j$-th channel of the nonlinear spatial-spectral feature map $F^{\Phi}$, $W_{j}$ is the $j$-th implicit linear convolution kernel that has a same size with image patches, and $\varphi$ is an implicit nonlinear function. This procedure is called as KPCA convolution. Note that when $\Phi$ is an identity map $\Phi\left(P\right)=P$, the Eq. (\ref{eq:22}) degenerates to
\begin{equation}
  F_{j}^{\Phi}=W_{j}*I, j=1,2, \cdots, k
\label{eq:23}
\end{equation}
which is the PCA convolution, thus the linear PCA convolution is a special case of KPCA convolution. 

\begin{figure}[t]

  \centering
  \includegraphics[scale=0.75]{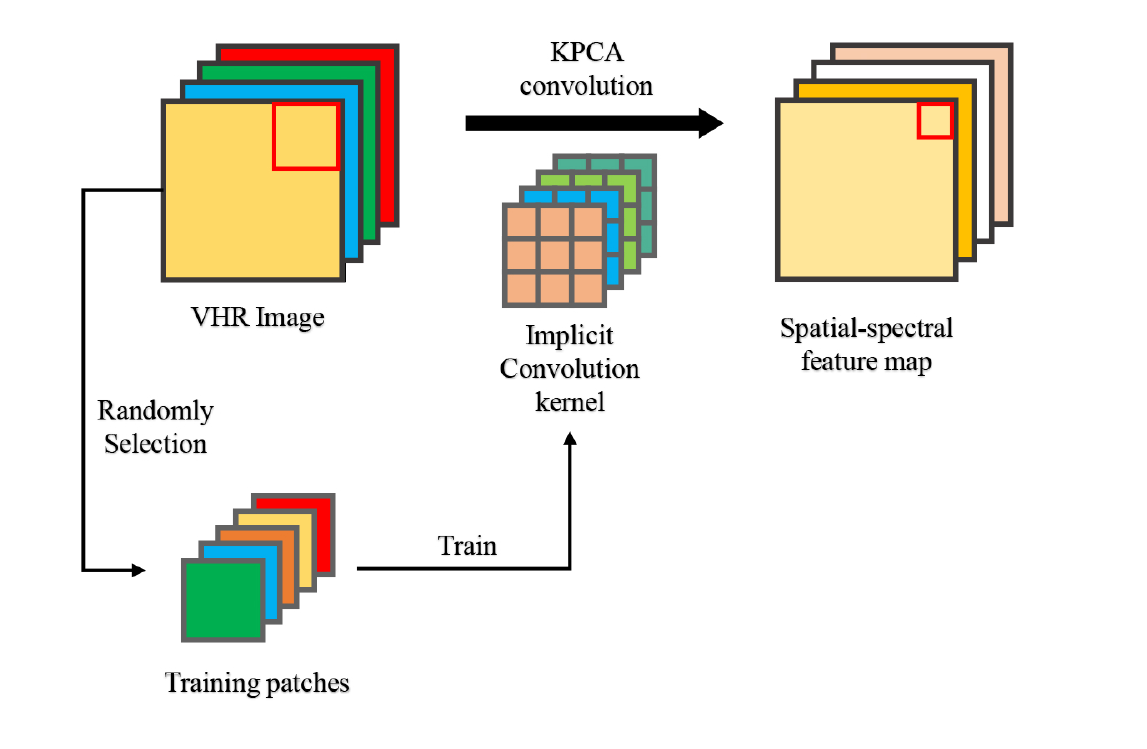}
  \caption{Illustration of feature extraction mode of KPCA convolution.}
  \label{fig:KPCAConv}
\end{figure}

\par As shown in Fig. \ref{fig:KPCAConv}, similar to the convolutional layer of CNN, the KPCA convolutional layer is able to extract representative spatial-spectral features from VHR images in a 2-D manner, but the learning process of parameters is totally unsupervised and does not require annotated data. The whole process of KPCA is summarized in Algorithm \ref{alg:KPCAConv}.

\begin{figure*}[ht]

  \centering
  \includegraphics[scale=0.3]{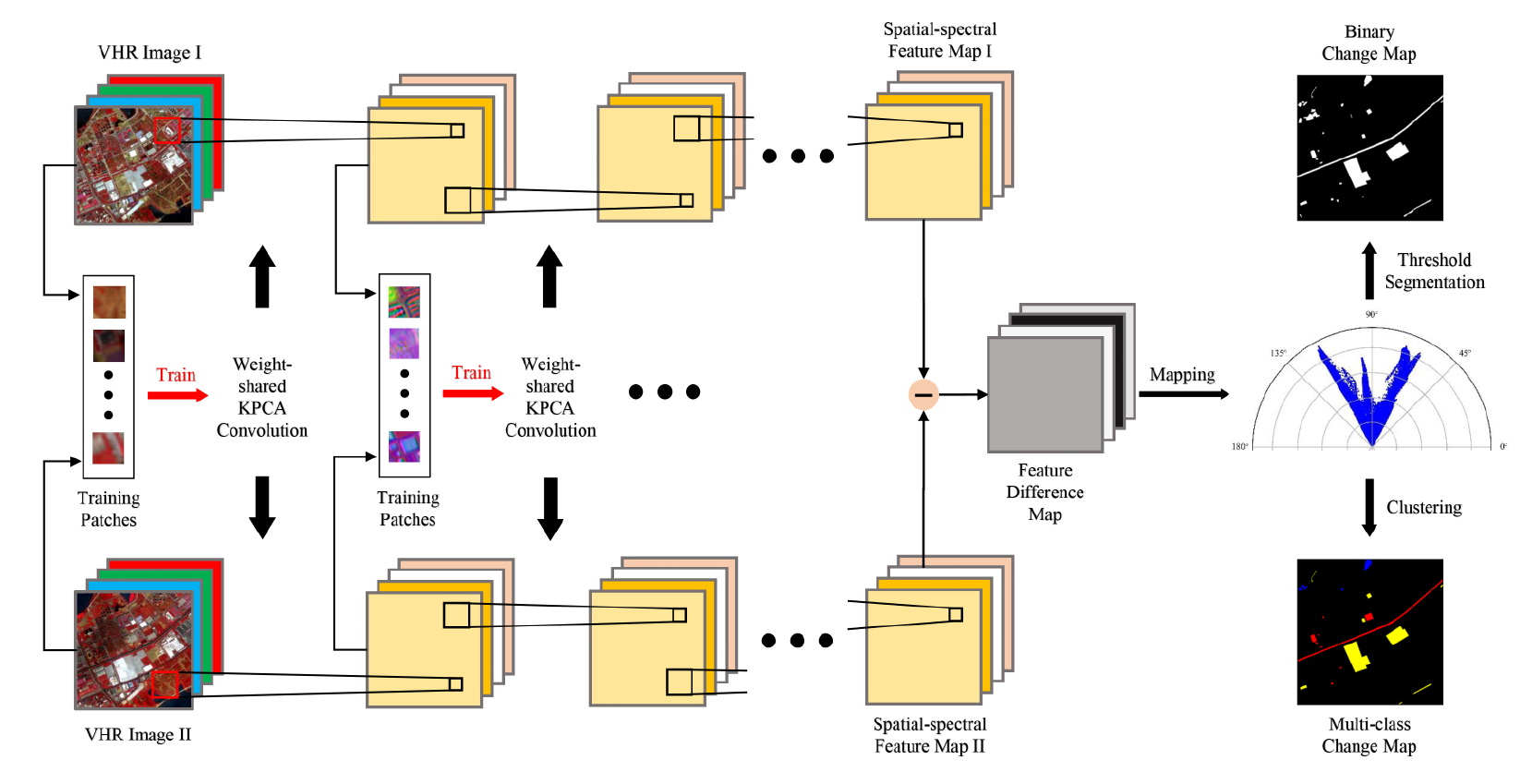}
  \caption{Overview of the CD architecture based on the proposed KPCA-MNet.}
  \label{fig:KPCAMNet}
\end{figure*}

\begin{algorithm}[htb]
	\caption{Process of Training and feature extraction for KPCA convolution.}
	\label{alg:KPCAConv}
	\begin{algorithmic}[1]
		\REQUIRE~~\\
		Input VHR image $I$;\\
		\ENSURE ~~\\
		Nonlinear spatial-spectral feature map $F^{\Phi}$;\\
		\STATE Generate vectorized image patches $P$ from $I$;
		\STATE Randomly select $n$ image patches from $P$ as training samples;
    \STATE Construct original kernel matrix $K$ with training samples;
    \STATE Calculate centralized kernel matrix $\tilde{K}$:\\ $$\tilde{K}_{ij}=\left(K-\frac{1}{M}1K-\frac{1}{M}K1+\frac{1}{M^2}1K1\right)_{ij};$$
    \STATE Solve KPCA problem to obtain coefficient $\alpha$;
    \STATE Normalize $\alpha$;
    \STATE Calculate the $j$-th channel of spatial-spectral feature map:\\ $$F_{j}^{\Phi}=\left(v_{j}^{\Phi}\right)^{T} \Phi(P)=\sum_{i=1}^{n} \alpha_{i}^{j}k^{\Phi}\left(p^{train}_{i},P\right);$$
    \STATE Stack each spatial-spectral feature map in channel to get $F_{j}^{\Phi}$;
    \RETURN $F_{j}^{\Phi}$;
	\end{algorithmic}
\end{algorithm}

\subsection{Deep Siamese KPCA Convolutional Mapping Network}
\par Based on the proposed KPCA convolution, the KPCA-MNet is designed for CD in multi-temporal VHR images. The architecture of the proposed KPCA-MNet is shown in Fig. \ref{fig:KPCAMNet}. First, the multi-temporal VHR images are pre-processed. Then, representative spatial-spectral feature maps are extracted from multi-temporal VHR images by several weight-shared KPCA convolutional layers. Next, by means of pixel-wise subtraction, the feature difference map is generated. Finally, the high-level difference features are mapped into a 2-D polar domain to employ change information efficiently and the final binary and multi-class CD results are generated by threshold segmentation methods and clustering algorithms.

\subsubsection{Data Pre-processing}
\par Before utilizing the proposed KPCA-MNet to detect changes, the first step is data pre-processing, including image registration and radiometric correction. Image registration is the process of aligning two or more images covering the same scene obtained at different times \cite{Zitova2003}. Given two multi-temporal images, only when the same pixels correspond to the same geographical location, can the CD between the corresponding areas be meaningful. There are three main steps in image registration: collecting matched point-pairs, establishing a transformation model, and transforming images. By these three steps, the two given multi-temporal images are geometrically aligned. 

\par Next, radiometric correction is performed to eliminate the radiometric difference between multi-temporal images caused by different imaging conditions, including sun angle, light intensity, atmospheric conditions, and so on. The specific method is radiometric relative normalization based on the z-score method, which standardize images with zero mean and unit variance. Given a multi-temporal spectral vector pair, denoted by $x_{i}=\left[x_{i}^{1},x_{i}^{2},\cdots,x_{i}^{N}\right]$ and $y_{i}=\left[y_{i}^{1},y_{i}^{2},\cdots,y_{i}^{N}\right]$, where $i$ indicates the pixel number and $N$ means the dimension of the spectral band. The radiometric relative normalization can be expressed as
\begin{equation}
  \hat{x}_{i}^{n}=\frac{x_{i}^{n}-\mu_{x_{n}}}{\sigma_{x_{n}}}\quad and \quad\hat{y}_{i}^{n}=\frac{y_{i}^{n}-\mu_{y_{n}}}{\sigma_{y_{n}}}
\label{eq:24}
\end{equation}
where $\mu_{x_{n}}$ is the mean and $\sigma_{x_{n}}$ is the variance for band $n$ of image $X$. Through radiometric correction, the radiometric difference between multi-temporal VHR images caused by different conditions would be suppressed.

\subsubsection{Deep KPCA Convolution}
\par After data pre-processing, we could train a KPCA convolutional layer, denoted as KC, to extracted nonlinear features from VHR images:
\begin{equation}
  F=KC\left(I\right)=\varphi\left(W*I\right).
\label{eq:25}
\end{equation}

\par Though the single KPCA convolutional layer could extract representative spatial-spectral features to some extent, it still has some inadequacies. First, the single KPCA convolutional layer has a fixed and limited receptive field and could only extract single scale features. However, there exist several scale features in VHR images. Besides, only a single layer means that it may not extract high-level features from VHR images. 

\par Therefore, for the purpose of extracting features from VHR images better, several KPCA convolutional layers are stacked for constructing a deep KPCA convolutional network to extract more representative features. Considering that CD involves two images and the processed VHR images in this paper are homogeneous, two siamese subnetworks $S_{1}$ and $S_{2}$ are designed to extract spatial-spectral features from two VHR images in parallel. Given two multi-temporal VHR images $I^{t_{1}}$ and $I^{t_{2}}$, without loss of generality, the outputs of the two subnetworks are
\begin{equation}
  \left\{
    \begin{aligned}
      &F^{t_{1}}=S_{1}\left(I^{t_{1}}\right)=K C_{1}^{L}\left(K C_{1}^{L-1} \cdots K C_{1}^{1}\left(I^{t_{1}}\right)\right)  \\
      &F^{t_{2}}=S_{2}\left(I^{t_{2}}\right)=K C_{2}^{L}\left(K C_{2}^{L-1} \cdots K C_{2}^{1}\left(I^{t_{2}}\right)\right)
    \end{aligned}
  \right.  
  \label{eq:26}    
\end{equation}

\par where $F^{t_{1}}$ and $F^{t_{2}}$ are the outputs, $KC^{i}_{1}$ is the $i$-th KPCA convolutional layer of $S_{1}$, and $L$ is the number of KPCA convolutional layer. Due to $S_{1}\equiv S_{2}$, the convolution kernel size, kernel parameters, and number of KPCA convolutional layers of $S_{1}$ and $S_{2}$ are all the same. Therefore, the two subnetworks extract features from multi-temporal VHR images via the exact same way. 

\par To train each KPCA convolutional layer, an unsupervised layer-wise training way is proposed. For the first layer, multiple training patch pairs are randomly selected to train the weight-shared KPCA convolutional layer. Each training patch pair is two multi-temporal image patches with the same size covering the same area. Next, multiple training patch pairs are selected from the two feature maps extracted by the first layer to train the second layer. By repeating this training-extracting process, each KPCA convolutional layer of KPCA-MNet can be trained and two representative spatial-spectral feature maps are generated. 

\subsubsection{Feature Mapping and Change Detection}
\par After two high-level spatial-spectral feature maps $F^{t_{1}},F^{t_{2}}\in R^{h\times w\times c_{L}}$ are extracted by subnetworks, the pixel-wise subtraction operation is implemented to get the feature difference map $D$, which contains abundant change information. To comprehensively utilize these change information, the feature magnitude (FM) $\rho$ is calculated as
\begin{equation}
  \rho=\sqrt{\sum_{i=1}^{C_{L}} D_{i}^{2}}=\sqrt{\sum_{i=1}^{C_{L}}\left(F_{i}^{t}-F_{i}^{t_{2}}\right)^{2}}, \rho \in\left[0, \rho_{\max }\right],
  \label{eq:27}    
\end{equation}
where $c_{L}$ is the dimension of channel, $D_{i}$ is the $i$-th channel of the feature difference map. $\rho$ carries information about whether change exists. The larger $\rho_{i}$ is, the higher change probability of pixel $i$ is. Through a threshold segmentation method, such as OTSU \cite{otsu1979} and EM \cite{moon1996}, the binary CD result can be generated. 

\begin{figure}[t]

  \centering
  \includegraphics[scale=1.3]{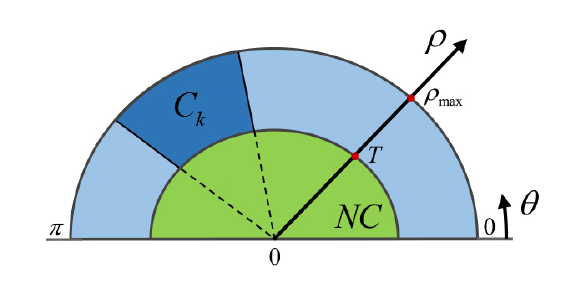}
  \caption{The 2-D polar domain of KPCA-MNet based on $\rho$ and $\theta$, where $NC$ indicates non-change, $C_{k}$ represents the $k$-th change type and $T$ means the threshold of segmentation.}
  \label{fig:2DPolar}
\end{figure}

\par At present, most of the unsupervised CD methods stop here, namely obtaining binary change maps. Nevertheless, the representative high-level spatial-spectral feature maps can be generated by the proposed method, in which the different objects could be more discriminative. That means the feature difference map is able to contain abundant information of different kind of change, which makes unsupervised multi-class CD possible. 

\par An intuitive idea is directly performing clustering algorithm on the feature difference map to get the multi-class change map. But plenty of information in feature difference map is redundant. Besides, the amount of information contained in different channels of the feature difference map is also different. Thus, performing clustering algorithms in the original difference space is often not effective to get the right change class \cite{lee2007,Saha2019} (this conclusion is also verified in section V-D). Inspired by the change vector direction measure proposed in \cite{kruse1993,Bovolo2012}, a weighting feature direction (WFD) is developed that maps the high-level change information to a 1-D variable
\begin{equation}
  \theta=\arccos \left[\left(\sum_{i=1}^{c_{L}} \lambda_{i}^{\Phi} D_{i}\right) / \sqrt{\sum_{i=1}^{c_{L}}\left(\lambda_{i}^{\Phi}\right)^{2} \sum_{i=1}^{c_{L}} D_{i}^{2}}\right], \theta \in[0, \pi],  
  \label{eq:28}    
\end{equation}
where $\lambda^{\Phi}_{i}$ is the eigenvalue corresponding to the $i$-th convolution kernel of the last KPCA convolutional layer in KPCANet. Since the larger eigenvalue means more information, in Eq. (\ref{eq:28}), the channel of feature difference map corresponding to the larger eigenvalue would play a more important role in calculating $\theta$. Despite existing some loss of information, the different kinds of change information are effectively utilized and mapped into $\theta$. By performing clustering algorithm on $\theta$, the multi-class CD results could be generated. 

\par Eventually, as shown in Fig. \ref{fig:2DPolar}, the high-level difference features are mapped into a 2-D polar domain based on $\rho$ and $\theta$. If only the binary change map is required, a threshold segmentation method is adopted to get the result according to $\rho$. If the aim is a multi-class CD map, a threshold segmentation method is first implemented to distinguish change pixels from unchanged ones. After that, a clustering algorithm is performing on $\theta$ to classify each changed pixel into specific change kind. 

\par Summarizing all of the aforementioned contents, the whole detailed process of KPCA-MNet is elaborated in Algorithm \ref{alg:KPCAMNet}.

\begin{algorithm}[htb]
	\caption{Process of Change Detection for KPCA-MNet.}
	\label{alg:KPCAMNet}
	\begin{algorithmic}[1]
		\REQUIRE~~\\
		Multi-temporal input VHR Image $I^{t_{1}}$ and $I^{t_{2}}$;\\
		\ENSURE ~~\\
		The binary change map $C_{B}$ and multi-class change detection map $C_{M}$;\\
		\STATE Align $I^{t_{1}}$ and $I^{t_{2}}$ geometrically;
    \STATE Standardize $I^{t_{1}}$ and $I^{t_{2}}$ using z-score method;
    \WHILE {$i < network\_depth$}
    \STATE Generate image patches or feature patches $P^{t_{1}}$ and $P^{t_{2}}$;
    \STATE Randomly select $n$ patch-pairs from $P^{t_{1}}$ and $P^{t_{2}}$ as training samples $P_{train}^{t}$;
    \STATE Train $i$-th weight-shared KPCA convolutional layer $KC^{i}$;
    \STATE Extract spatial-spectral features $F^{t_{1}}$ and $F^{t_{2}}$;
    \STATE $i$++;
    \ENDWHILE
    \STATE Calculate feature difference map $D$: \\ $$D=F^{t_{1}}-F^{t_{2}};$$
    \STATE Calculate the feature magnitude $\rho$:\\ $$\rho=\sqrt{\sum_{i=1}^{B}D^{2}_{i}};$$
    \STATE Calculate the weighting feature direction $\theta$:\\ $$\theta=\arccos \left[\left(\sum_{i=1}^{c_{L}} \lambda_{i}^{\Phi} D_{i}\right) / \sqrt{\sum_{i=1}^{c_{L}}\left(\lambda_{i}^{\Phi}\right)^{2} \sum_{i=1}^{c_{L}} D_{i}^{2}}\right];$$
    \STATE Perform threshold segmentation to get the binary change map $C_{B}$;
    \STATE Implement clustering algorithm to get the multi-class change map $C_{M}$;
    \RETURN $C_{B}$ and $C_{M}$;
	\end{algorithmic}
\end{algorithm}

\begin{figure}[ht]
  \centering

  \subfloat[]{
    \includegraphics[width=1.05in]{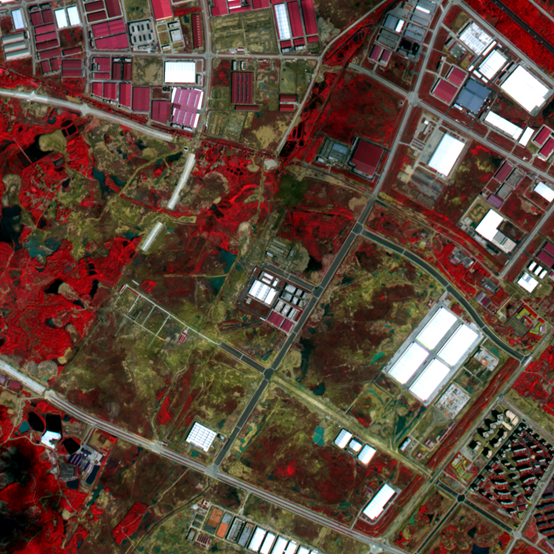}
  \label{fig_first_case}}
  \hfil
  \subfloat[]{
    \includegraphics[width=1.05in]{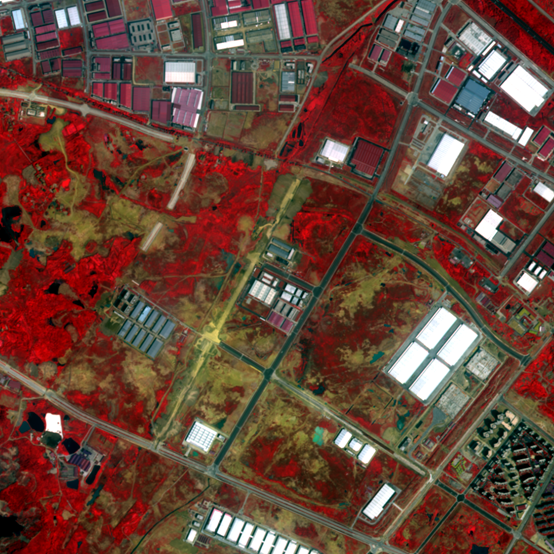}
  \label{fig_second_case}}
  \hfil
  \subfloat[]{
    \includegraphics[width=1.05in]{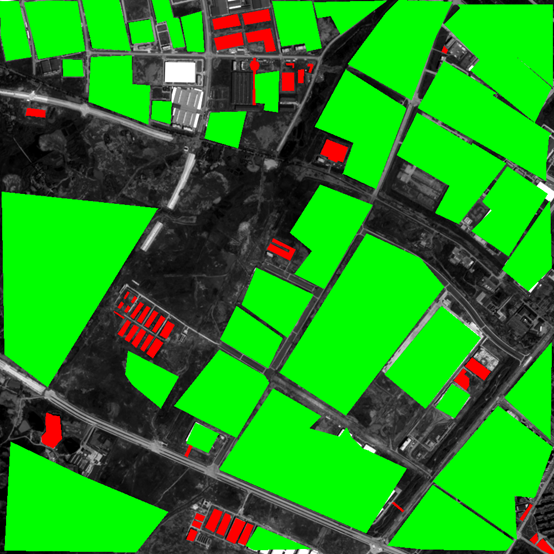}
  \label{fig_third_case}}

  \caption{The pseudo-color VHR images of (a) WH-1, (b) WH-2. (c) is ground truth, where red indicates changed region and green indicates unchanged region.}
  \label{WH_dataset}
\end{figure}

\begin{figure}[ht]
  \centering

  \subfloat[]{
    \includegraphics[width=1.05in]{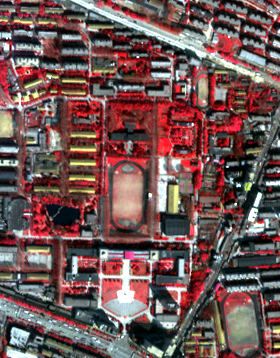}
  \label{fig_first_case}}
  \hfil
  \subfloat[]{
    \includegraphics[width=1.05in]{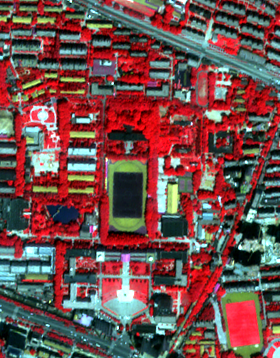}
  \label{fig_second_case}}
  \hfil
  \subfloat[]{
    \includegraphics[width=1.05in]{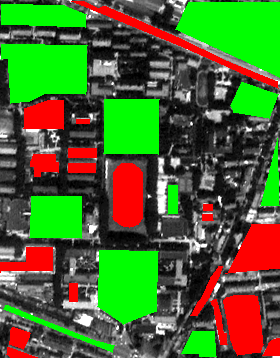}
  \label{fig_third_case}}

  \caption{The pseudo-color VHR images of (a) QU-1, (b) QU-2. (c) is ground truth, where red indicates changed region and green indicates unchanged region.}
  \label{QU_dataset}
\end{figure}

\section{Experiment of Binary Change Detection}\label{sec:4}

\subsection{Data Description}
\par In the experiment of binary change detection, two VHR images are adopted to evaluate the proposed method. The first data set was acquired by GF-2 on April 4, 2016 and September 1, 2016, covering the city of Wuhan, China, denoted as WH. The multi-temporal VHR image pairs in this data set consist of 4 spectral bands, i.e. red band, green band, blue band and near-infrared band, with 1000 $\times$ 1000 pixels and they have a spatial resolution of 4 m. The pseudo-color VHR images and the corresponding ground truth of changes and non-changes are shown in Fig. \ref{WH_dataset}. In the ground truth, red indicates changed regions, green indicates unchanged regions, and the remaining pixels are undefined. In the WH data set, a small part of buildings suffers from an “over-exposed” problem, which breaks the linear relationship of the digital numbers of unchanged regions between multi-temporal images and cannot be eliminated by radiometric normalization. Therefore, the “over-exposed” problem makes accurate CD more difficult.

\par The second VHR data set was gathered by QuickBird on 2002 and 2005 over the Wuhan University, denoted as QU. The multi-temporal VHR image pairs in the QU data set consist of 4 spectral bands with 358 $\times$ 280 pixels and they have a spatial resolution of 2.4 m. The pseudo-color VHR images and the ground truth are shown in Fig. \ref{QU_dataset}. In the ground truth, red indicates changed regions, green indicates unchanged regions, and the remaining pixels are undefined.

\begin{figure*}[ht]
  \centering
  \subfloat[]{
    \includegraphics[width=1.05in]{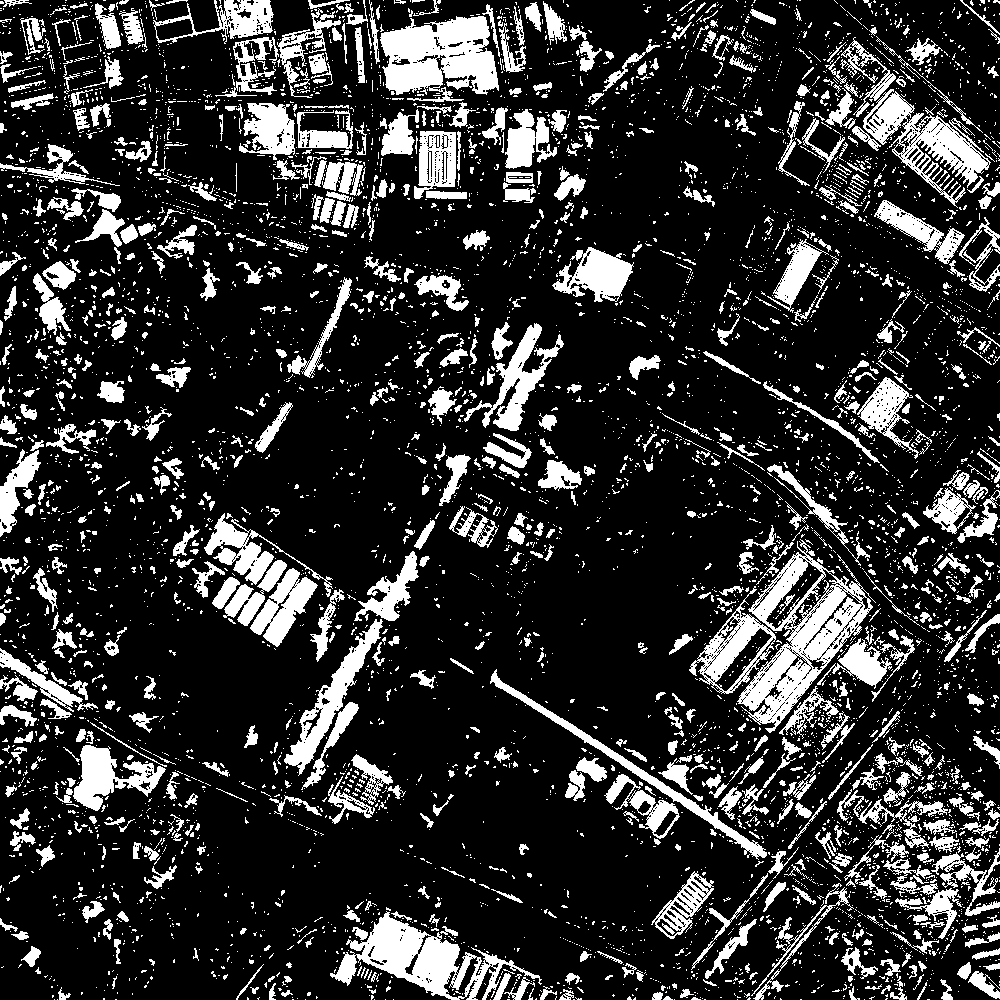}
  \label{WH_MAD}}
  \hfil
  \subfloat[]{
    \includegraphics[width=1.05in]{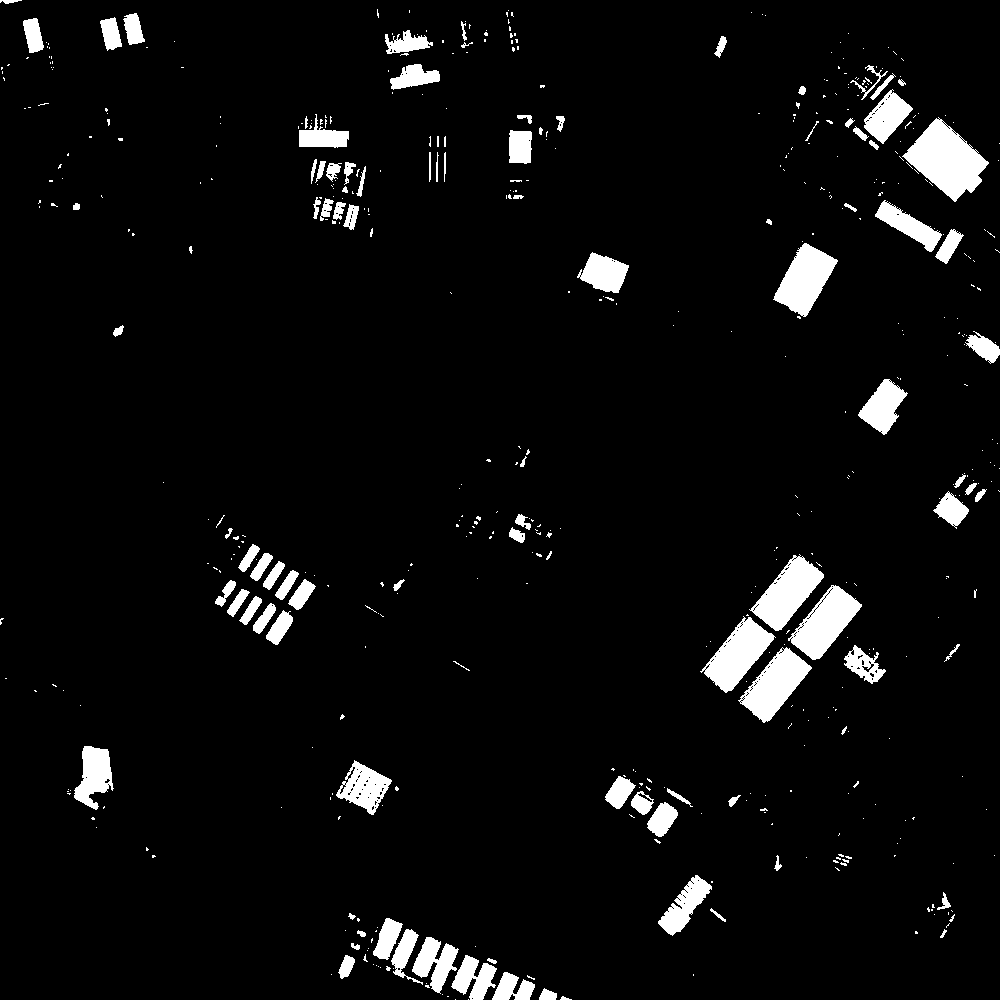}
  \label{WH_IRMAD}}
  \hfil
  \subfloat[]{
    \includegraphics[width=1.05in]{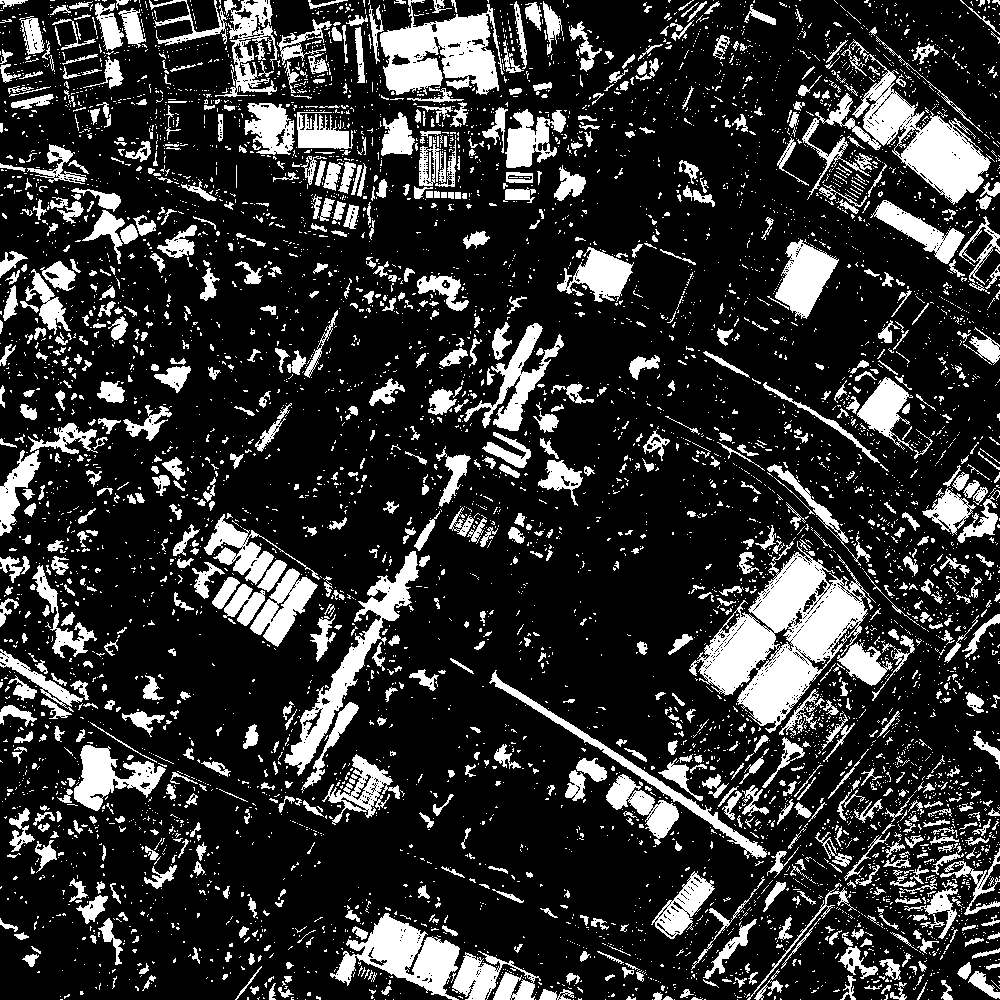}
  \label{WH_USFA}}
  \hfil
  \subfloat[]{
    \includegraphics[width=1.05in]{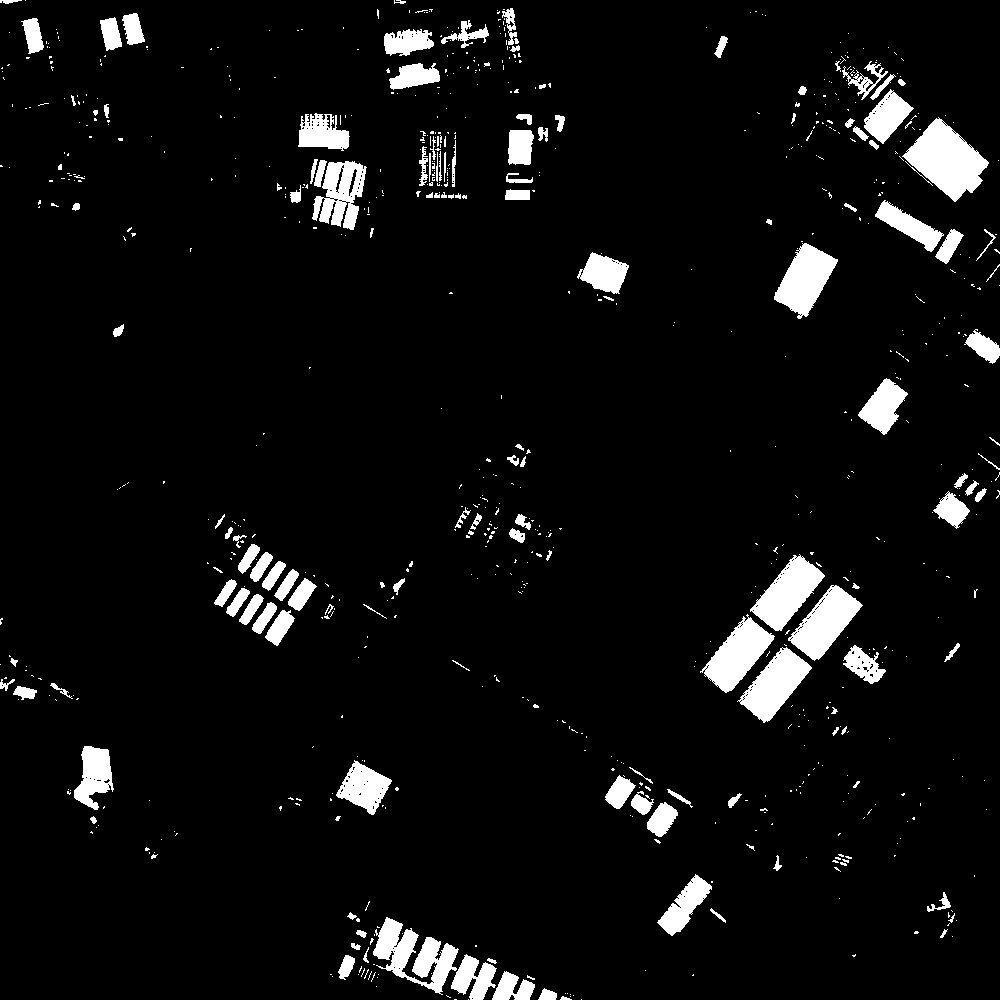}
  \label{WH_ISFA}}
  \hfil
  \subfloat[]{
    \includegraphics[width=1.05in]{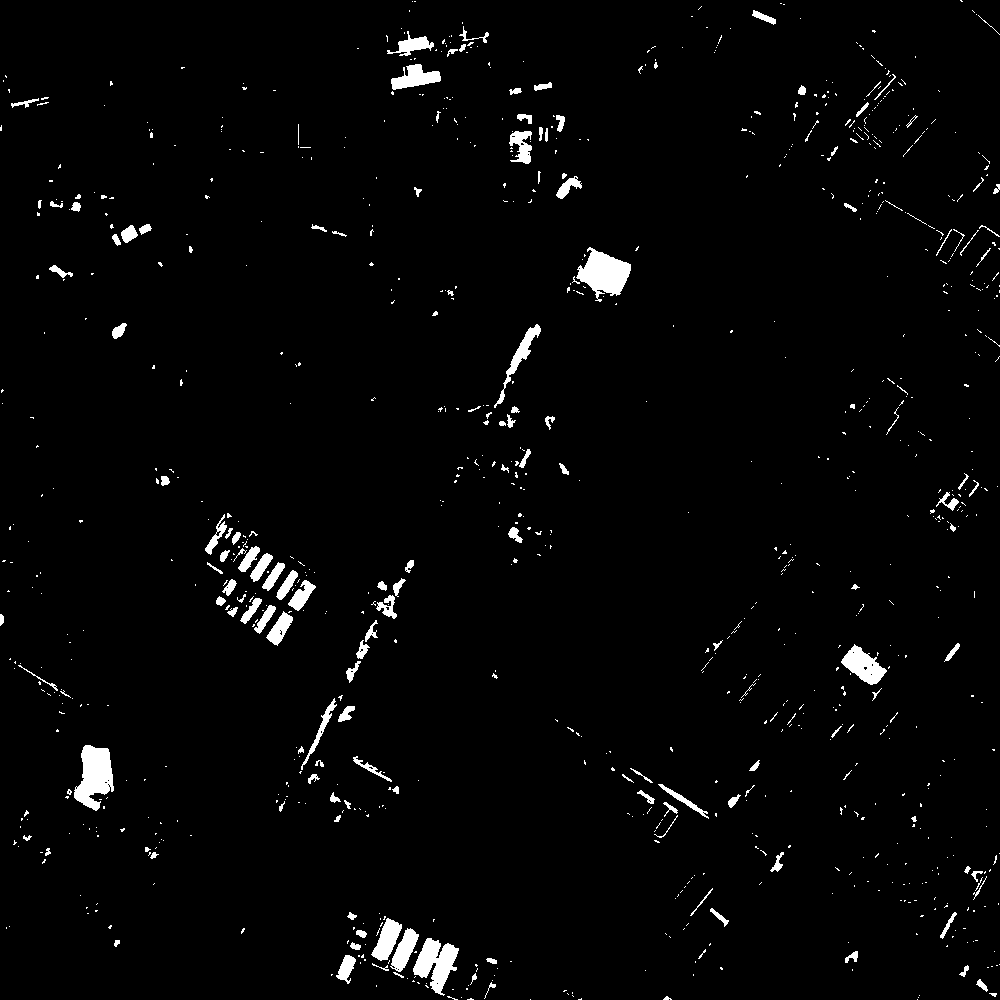}
  \label{WH_CVA}}
  \hfil
  \subfloat[]{
    \includegraphics[width=1.05in]{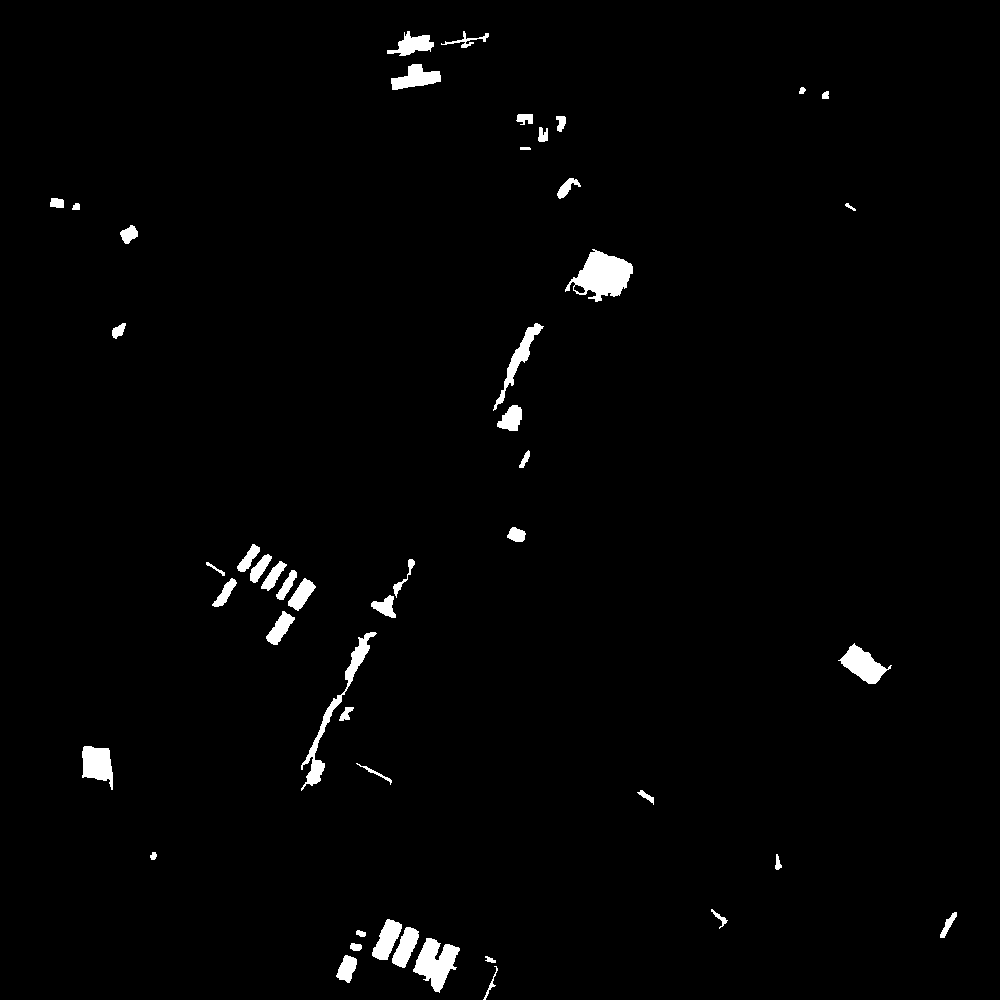}
  \label{WH_OBCD}}

  \subfloat[]{
    \includegraphics[width=1.05in]{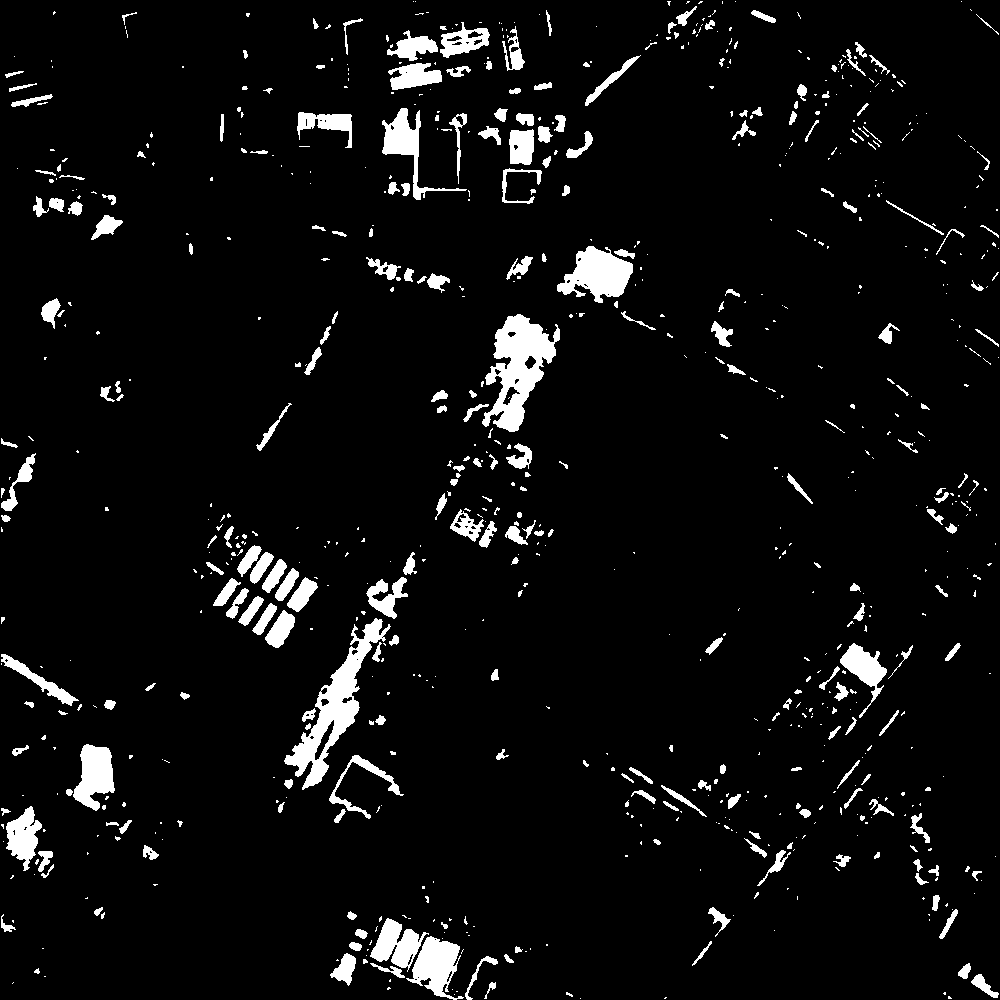}
  \label{WH_PCA-Kmeans}}
  \hfil
  \subfloat[]{
    \includegraphics[width=1.05in]{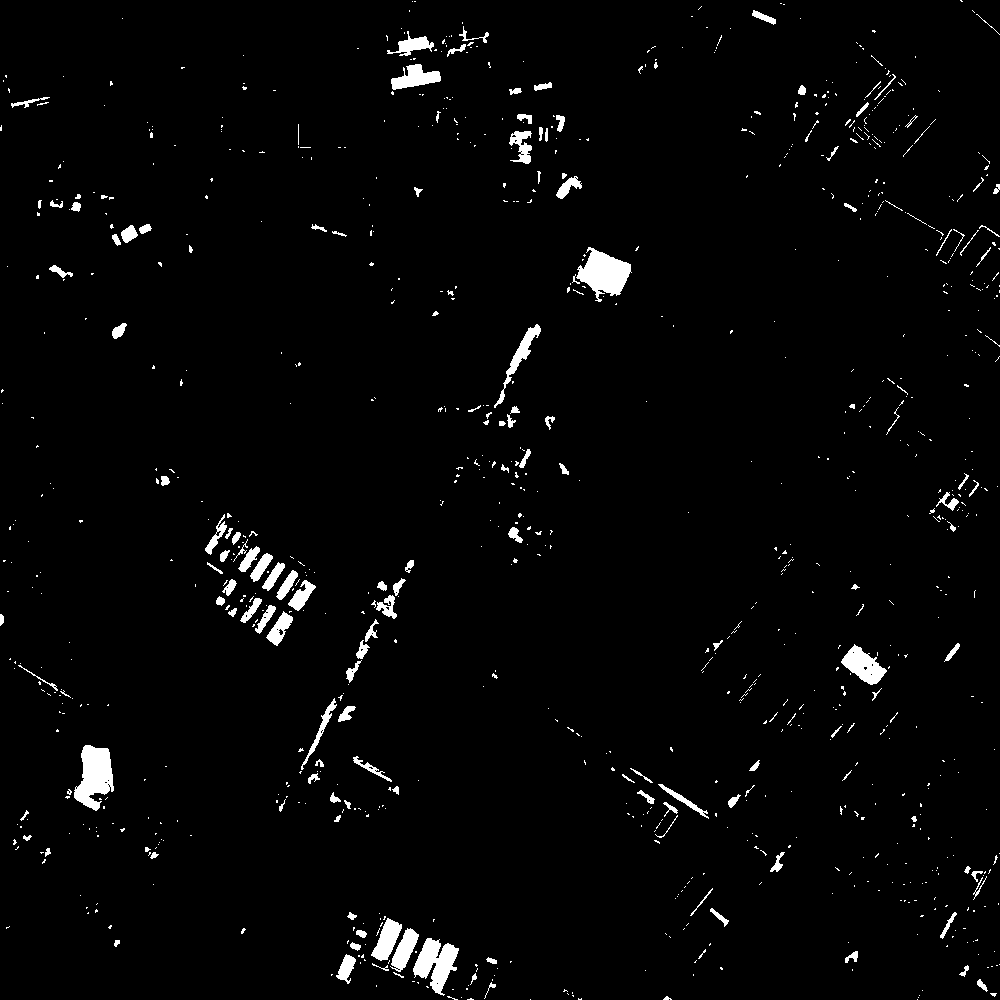}
  \label{WH_SVM}}
  \hfil
  \subfloat[]{
    \includegraphics[width=1.05in]{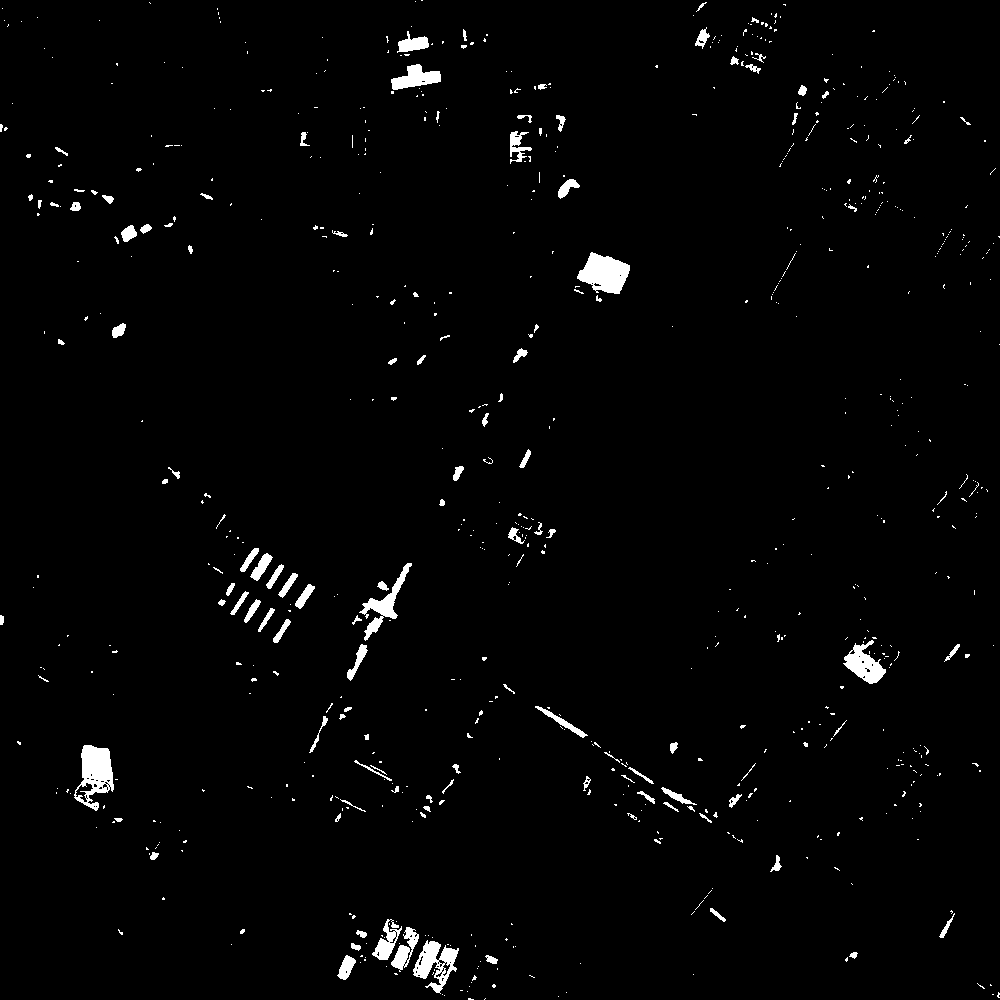}
  \label{WH_DSCN}}
  \hfil
  \subfloat[]{
    \includegraphics[width=1.05in]{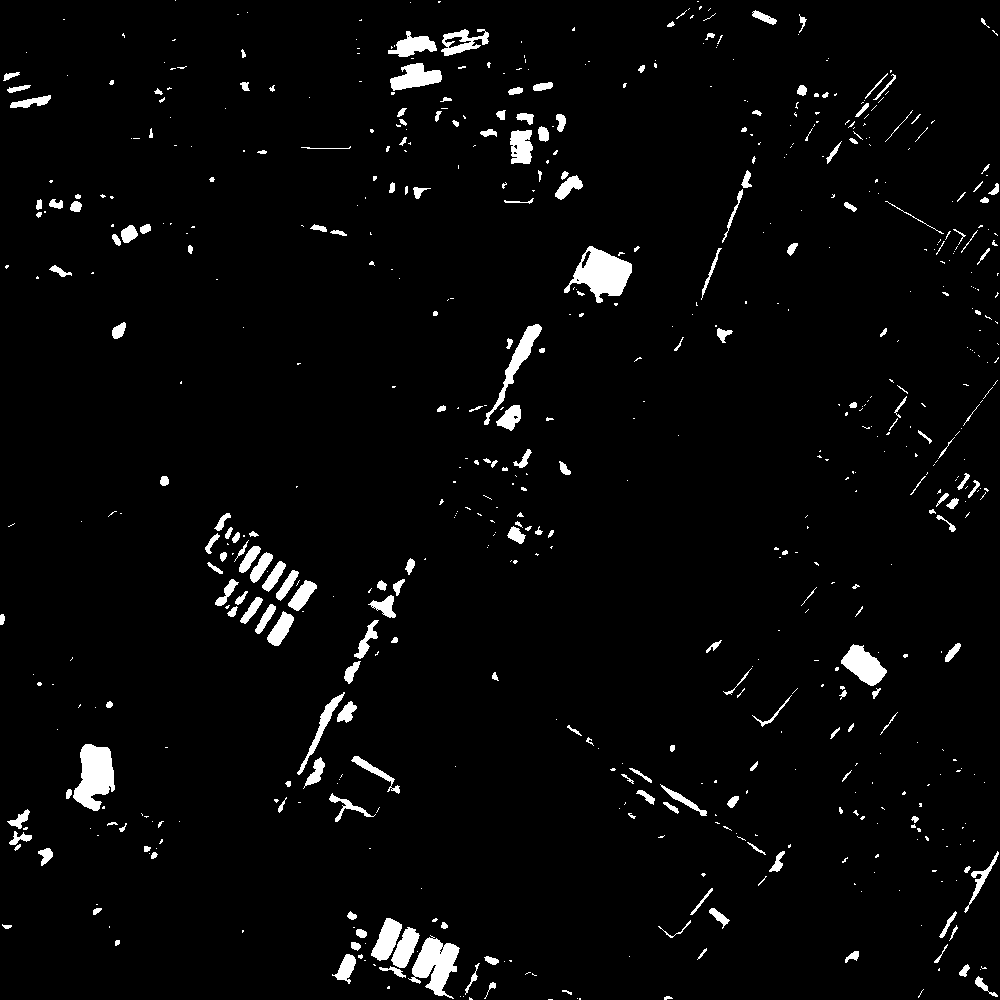}
  \label{WH_RNN}}
  \hfil
  \subfloat[]{
    \includegraphics[width=1.05in]{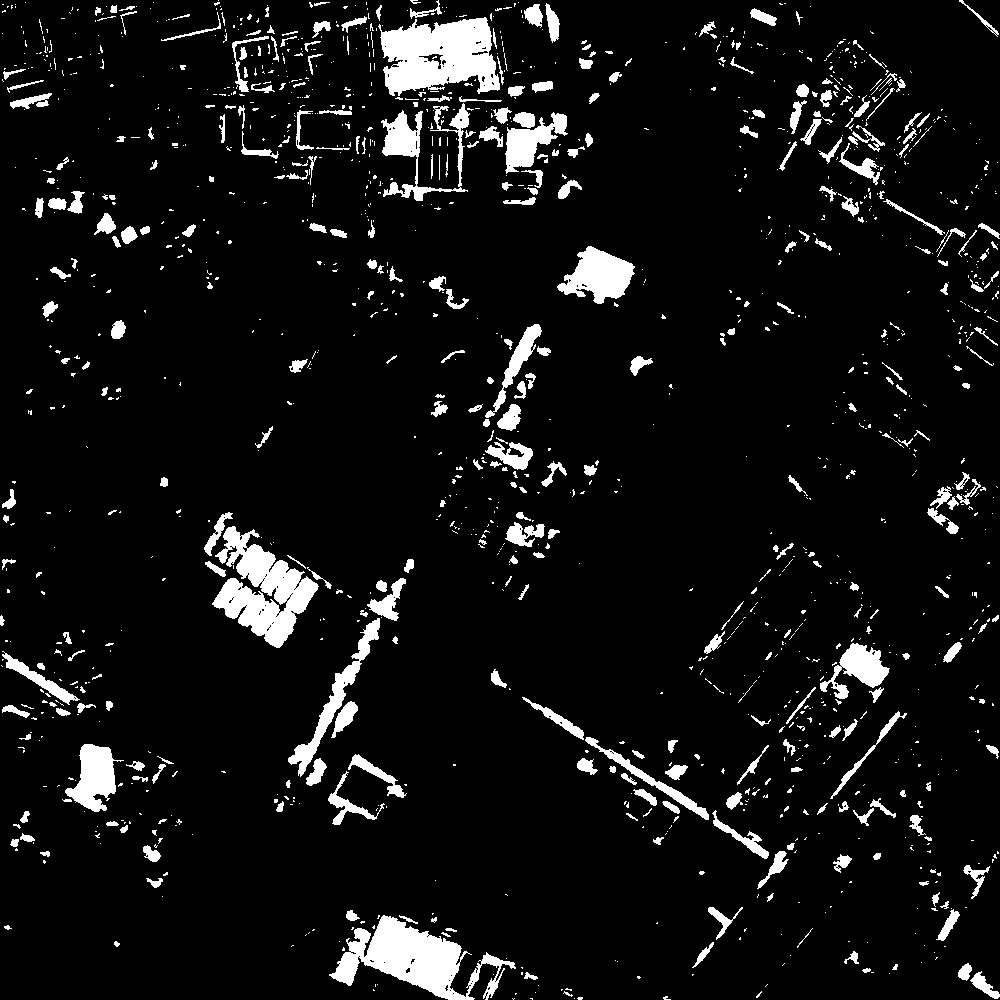}
  \label{WH_SAE}}
  \hfil
  \subfloat[]{
    \includegraphics[width=1.05in]{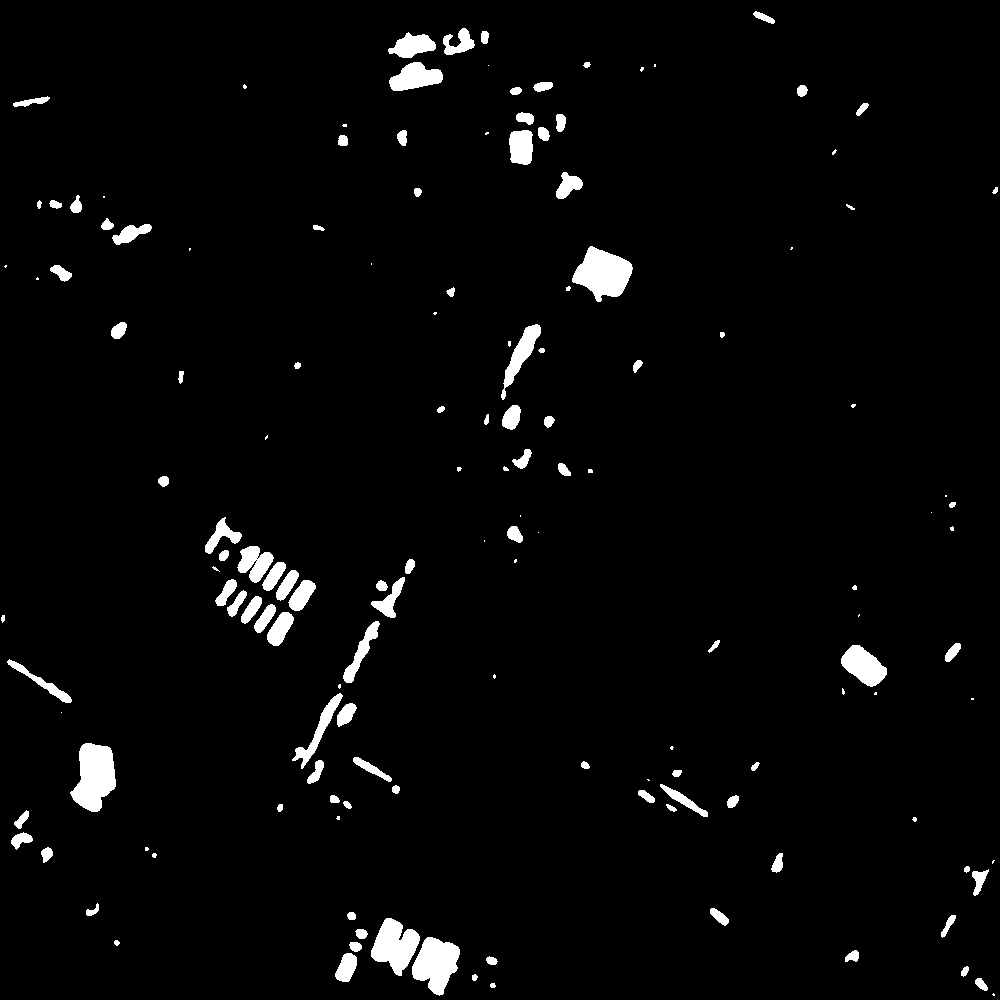}
  \label{WH_KPCAMNet}}
  \caption{Binary change maps obtained by the proposed method and comparison methods on the WH data set. (a) MAD. (b) IRMAD. (c) USFA. (d) ISFA. (e) CVA. (f) OBCD. (g) PCA-Kmeans. (h) SVM. (i) DSCN. (j) RNN-CD. (k) SAE. (l) KPCA-MNet.}
  \label{WH_result}
\end{figure*}

\subsection{Experiment settings}
\par Several hyper-parameters influence the performance of the proposed KPCA-MNet model, including the selection of kernel function $k\left(x, y\right)$, the number of training patches $n$, the number and size of KPCA convolution kernel, $p$ and $w$, and the network depth $L$. For simplicity, the hyper-parameters of different KPCA convolutional layers use the same values. The choices of these parameters in the experiment are listed in Table I. And the specific influence of these parameters at different values is discussed in section IV-D.

\par To validate the performance of the proposed KPCA-MNet, it is compared with the most widely used CD methods, including CVA \cite{Sharma2007}, MAD \cite{Nielsen1997}, IRMAD \cite{Nielsen2007}, USFA \cite{Wu2014}, ISFA \cite{Wu2014}, PCA-Kmeans \cite{Celik2009}, OBCD \cite{Desclee2006}, SVM, DSCN \cite{Zhan2017}, RNN-CD \cite{Lyu2016} and SAE \cite{Bengio2007}. MAD is a CD method based on the canonical correlation analysis (CCA), which is first proposed in \cite{Nielsen1997}. Based on the SFA theory, USFA computes a projecting matrix to suppress pseudo-change pixels and highlight changed pixels. IRMAD and ISFA are iterative versions of MAD and USFA. DSCN is a deep siamese convolutional network and performs well in aerial images. RNN-CD is an RNN-based CD network proposed in \cite{Lyu2016}, which has shown good performance in CD. SAE is an unsupervised deep learning method constructed by stacking several autoencoders. By layer-wise greedy training, SAE can extract essential features of input data. Among these methods, SVM, DSCN, and RNN-CD are supervised methods. For the sake of fairness, a common way is that selecting the annotated training samples by the unsupervised automatic pre-detection method \cite{Gao2016,Li2019a,Du2019a}. Therefore, though training these three methods is a supervised learning fashion, the whole process is a totally unsupervised process without any priori-knowledge. In OBCD, the mean and standard deviation of each band are utilized to generate objects. The RBF is chosen as the kernel function of SVM. The optimal hyper-parameters $C$ and $\gamma$ are chosen by grid search. The hyper-parameters of the remaining unsupervised methods use the optimal values recommended in their original references. K-means, OTSU, and FCM are adopted as threshold segmentation methods, the best results are chosen as the final results. 

\par The binary CD problem could be regarded as a binary classification problem for change and non-change. In the binary change map, white pixels denote changed areas and black pixels are unchanged areas. To comprehensively evaluate the performance of the proposed model and comparison methods, the following evaluation criteria are utilized in the accuracy assessment. 

\begin{enumerate} 
  \item False Positive (FP). FP is the number of unchanged pixels which are falsely detected as changed ones.
  \item False Negative (FN). FN is the number of pixels which are detected as unchanged pixels but actually changed ones.
  \item Overall Error (OE). OE is the total number of pixels that are misclassified. OE can be calculated by OE=FP+FN.
  \item Overall accuracy (OA). OA shows the number pixels that are detected correctly, divided by the number of total pixels. OA can be given by OE: OA=1 – OE / N, where N is the number of total pixels.
  \item Kappa Coefficient (KP). KP is a similarity measurement between the final CD results and ground truth. In binary CD, KP is defined as KP=(OA-PE) / (1-PE), where PE=((TP+FP)$\cdot$(TP+FN) +(TN+FN)$\cdot$(FP+TN)) / $N^{2}$.
\end{enumerate}

\subsection{Experimental Results and Analysis}
\par The first is the experimental results on the WH data set. The binary change maps obtained by the proposed KPCA-MNet and all comparison methods are illustrated in Fig. \ref{WH_result}. As shown in Fig. 6-(a) and Fig. 6-(c), the performance of MAD and SFA are unsatisfactory, numerous salt-and-pepper noise exists in the result of IRMAD and ISFA. Employing iterative reweighting, the results acquired by IRMAD and ISFA are relatively cleaner in visual. However, plenty of changed regions are miss detected. Besides, some unchanged buildings are misclassified as changed ones, which means IRMAD and ISFA are confused by the “over-exposed” problem. Compared with IRMAD and ISFA, the binary change map obtained by CVA is better. Nonetheless, since the pixels are processed in isolation and only spectral information is considered, some small objects, such as road, are not classified correctly and a lot of margins of buildings are falsely detected as change. Adopting objects to replace pixels, in OBCD, the object becomes the basic unit of CD. Therefore, in Fig. 6-(f), there is almost no salt-and-pepper noise in the results of OBCD. But due to only utilizing low-level features and limited to the performance of segmentation algorithm, lots of building changes are not detected. By means of transforming spatial blocks, PCA-Kmeans utilizes spatial context information to detect changes, thus a majority of building changes is detected and the result is not affected by the “over-exposed” problem. Nevertheless, some margins of building are also falsely detected.

\begin{figure*}[t]
  \centering
  \subfloat[]{
    \includegraphics[width=1.05in]{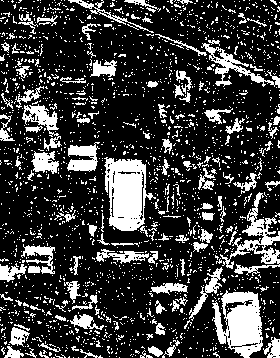}
  \label{QU_MAD}}
  \hfil
  \subfloat[]{
    \includegraphics[width=1.05in]{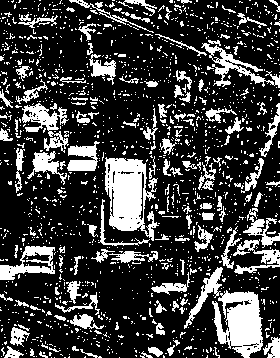}
  \label{QU_IRMAD}}
  \hfil
  \subfloat[]{
    \includegraphics[width=1.05in]{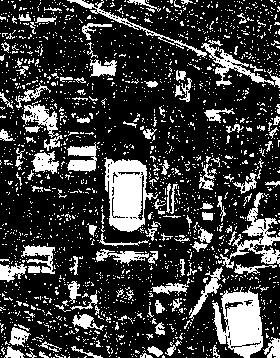}
  \label{QU_USFA}}
  \hfil
  \subfloat[]{
    \includegraphics[width=1.05in]{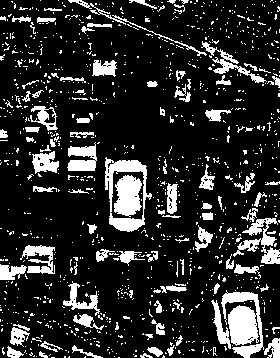}
  \label{QU_ISFA}}
  \hfil
  \subfloat[]{
    \includegraphics[width=1.05in]{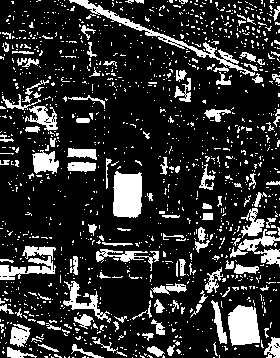}
  \label{QU_CVA}}
  \hfil
  \subfloat[]{
    \includegraphics[width=1.05in]{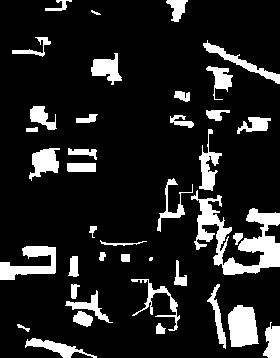}
  \label{QU_OBCD}}

  \subfloat[]{
    \includegraphics[width=1.05in]{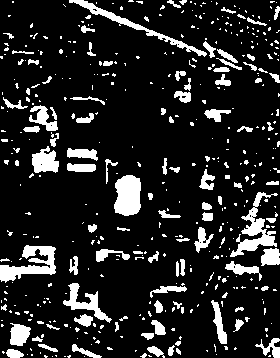}
  \label{QU_PCA-Kmeans}}
  \hfil
  \subfloat[]{
    \includegraphics[width=1.05in]{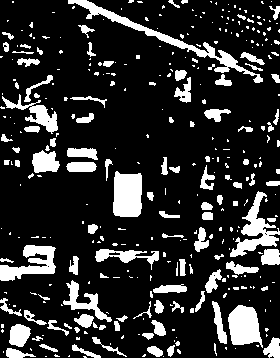}
  \label{QU_SVM}}
  \hfil
  \subfloat[]{
    \includegraphics[width=1.05in]{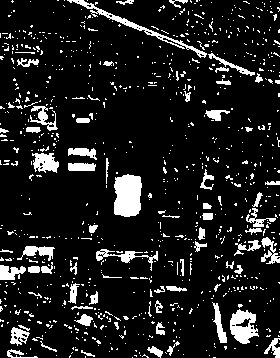}
  \label{QU_DSCN}}
  \hfil
  \subfloat[]{
    \includegraphics[width=1.05in]{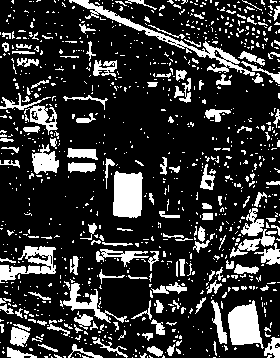}
  \label{QU_RNN}}
  \hfil
  \subfloat[]{
    \includegraphics[width=1.05in]{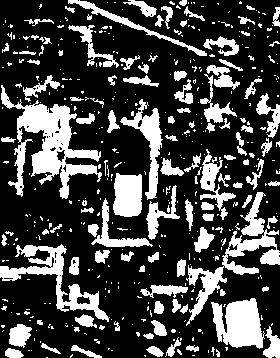}
  \label{QU_SAE}}
  \hfil
  \subfloat[]{
    \includegraphics[width=1.05in]{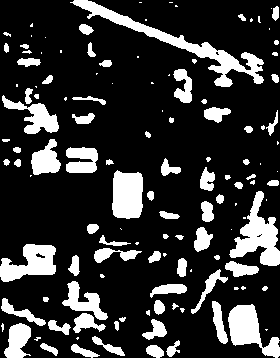}
  \label{QU_KPCAMNet}}
  \caption{Binary change maps obtained by the proposed method and comparison methods on the QU data set. (a) MAD. (b) IRMAD. (c) USFA. (d) ISFA. (e) CVA. (f) OBCD. (g) PCA-Kmeans. (h) SVM. (i) DSCN. (j) RNN-CD. (k) SAE. (l) KPCA-MNet.}
  \label{QU_result}
\end{figure*}

\begin{table}[t]
  \renewcommand{\arraystretch}{1.3}
  \caption{Hyper-parameter Setting For KPCA-MNet on Two Data Sets.}
  \label{HY_table}
  \centering
  \begin{tabular}{c c c c c c}
    \hline
    \bfseries Data set & \bfseries $k(x, y)$ & \bfseries $n$ & \bfseries $p$ & \bfseries $w$ & \bfseries $L$\\
    \hline\hline
    WH 	& RBF           & 200          & 8          & 5          & 3          \\								
    QU  & RBF           & 200          & 8          & 11          & 3          \\									
    \hline
  \end{tabular}
\end{table}

\par Limited to the relatively weak fitting ability of SVM model, in Fig. 6-(h), a small part of changed regions are broken inside and some groundworks before building-over are not detected. Utilizing a deep siamese convolution structure to extract spatial-spectral features and adopting a threshold segmentation algorithm to get the results, in the CD results generated by DSCN, margins of building are correctly classified. However, many changes in build-up regions are not detected. Adopting the RNN architecture to learn the universal rules of change, the change information between multi-temporal VHR images is captured by RNN-CD, and the acquired binary change map is more accurate than the ones acquired by SVM and DSCN. But RNN-CD cannot extract spatial features efficiently, which results in the misdetection of building margins and internal fragmentation of changed regions. Learning the essential features from image patches with several AE, SAE generates a good result, in which almost of changed pixels are classified correctly. The downside is that the fixed conceptive filed and fully-connected structure of SAE damage the feature extraction procedure to some degree. Hence, in Fig. 6-(k), there exists some noise and a part of unchanged pixels is falsely detected as changed ones. Besides, the results of SAE are affected by the “over-exposed” problem.

\begin{table}[t]
  \renewcommand{\arraystretch}{1.3}
  \caption{Accuracy assessment on binary change maps obtained by different methods on the WH data set}
  \label{WH_table}
  \centering
  \begin{tabular}{c c c c c c}
    \hline
    \bfseries Method & \bfseries FP & \bfseries FN & \bfseries OE & \bfseries OA & \bfseries KC\\
    \hline\hline
    MAD	& 79298	& \underline{2066}	& 78364	& 	0.8446	& 0.2662 \\ 				
    IRMAD	& 31154	& 8596	& 39750	& 	0.9212	& 0.3289  \\ 												
    USFA	& 86977		& 2271	& 89248	& 0.8230	& 0.2335	 \\ 												
    ISFA & 33155	& 7194	& 40349	& 0.9200	& 0.3530 \\ 				
    CVA	& 3432	& 8654	& 12086	& 0.9760	& 0.6409 \\ 					 							
    OBCD	& \textbf{617}	& 10201	& \underline{10818}	& \underline{0.9785}	& 0.6350 \\ 				
    PCA-Kmeans	& 10435	& 5401	& 15836	& 0.9686	& 0.6325 \\  									
    SVM	& 3556	& 8791	& 12347	& 0.9755	& 0.6330 \\ 	
    DSCN	& 2457	& 11046	& 13503	& 0.9732	& 0.5564 \\ 	
    RNN-CD	& 5426	& 7359	& 12785	& 0.9746	& 0.6515 \\
    SAE & 14869	& \textbf{1599}	& 16468	& 0.9673	& \underline{0.6750} \\
    KPCA-MNet	& 2525	& 6677	& \textbf{9202}	& \textbf{0.9819}	& \textbf{0.7316} \\ 	
    \hline
  \end{tabular}
\end{table}

\par Utilizing deep siamese KPCA convolutional architecture to extract representative high-level spatial-spectral features and adopting the feature magnitude to merge the change information in these features, the proposed KPCA-MNet obtains the best binary change map with less noise and more complete changed regions. What’s more, due to using spatial context information efficiently, KPCA-MNet is immune to the “over-exposed” problem. 

\par To better evaluate these methods, the quantitative results are listed in Table II. Similar to the conclusion derived from Fig. 6-(a) and Fig. 6-(c), MAD and USFA has a low KC of 0.2662 and 0.2335. And their iterative versions are slightly better, with a small OE and a higher KC. In fact, SFA and MAD are based on the central limit theorem, but the gaussianity of VHR images is not obvious, so it is difficult for these two methods to show good performance on VHR images. Due to the objects become the basic units in OBCD, thus the FP of OBCD is best, only 617 unchanged pixels are misclassified as changed ones. Trained on the samples selected by unsupervised automatic pre-detection, the accuracy of the three supervised methods are relatively good. Owing to the good capacity of feature extraction, SAE achieves the best FN. Finally, the proposed method outperforms the other compared methods with the best OE, OA, and KC. 

\begin{table}[t]
  \renewcommand{\arraystretch}{1.3}
  \caption{Accuracy assessment on binary change maps obtained by different methods on the QU data set}
  \label{WH_table}
  \centering
  \begin{tabular}{c c c c c c}
    \hline
    \bfseries Method & \bfseries FP & \bfseries FN & \bfseries OE & \bfseries OA & \bfseries KC\\
    \hline\hline
    MAD	& 3146	& 3611	& 6757	& 	0.8089	& 0.5640 \\ 				
    IRMAD	& 3100	& 3520	& 6620	& 	0.8128	& 0.5733  \\ 												
    USFA	& 3393		& \underline{3453}	& 6846	& 0.8064	& 0.5621	 \\ 												
    ISFA & 2021	& 4744	& 6765	& 0.8087	& 0.5409 \\ 				
    CVA	& 2595	& 4552	& 7147	& 0.7979	& 0.5233 \\ 					 							
    OBCD	& \textbf{709}	& 7009	& 7718 	& 0.7817	& 0.4293 \\ 				
    PCA-Kmeans	& 1371	& 6003	& 7374	& 0.7914	& 0.4766 \\  									
    SVM	& 2042	& 3932	& \underline{5974}	& \underline{0.8310}	& \underline{0.6022} \\ 	
    DSCN	& \underline{1361}	& 6363	& 7727	& 0.7815	& 0.4468 \\ 	
    RNN-CD	& 2929	& 4344	& 7273	& 0.7943	& 0.5208 \\
    SAE & 3670	& 3733	& 7403	& 0.7906	& 0.5265 \\
    KPCA-MNet	& 1673	& \textbf{1765}	& \textbf{3468}	& \textbf{0.9019}	& \textbf{0.7778} \\ 	
    \hline
  \end{tabular}
\end{table}

\par The CD results obtained by the proposed method and comparison methods on the QU data set are displayed in Fig. \ref{QU_result}. In the result acquired by MAD, a part of changed areas are falsely detected, and there exists plenty of salt-and-pepper noise. Induced by the complexity of VHR images, the problems in the binary change map of MAD are not solved by iterative reweighting operation, thus they also exist in the result of IRMAD. In Fig. 7-(c), the CD result of SFA is similar to the ones of MAD. Slightly different from MAD, after iterative reweighting, ISFA acquires the results with less noise. In the binary change maps of CVA and PCA-Kmeans (see in Fig. 7-(e) and Fig. 7-(g)), there are more changed pixels are misclassified. Besides, some unchanged buildings are falsely detected as changed ones. Similar to what we observed on the WH data set, there is almost no noise in the results generated by OBCD. However, lots of obvious changed regions are not recognized, such as the changes of the central playground. 
\begin{figure}[t]
  \centering

  \subfloat[]{
    \includegraphics[width=2.6in]{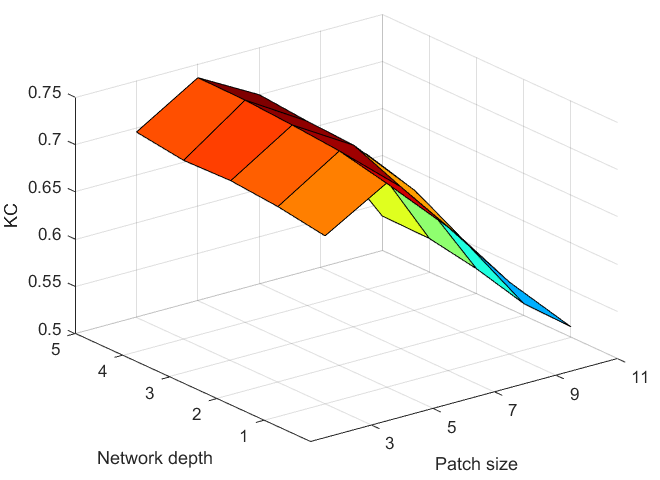}
  \label{fig_WH}}
  \hfil
  \subfloat[]{
    \includegraphics[width=2.6in]{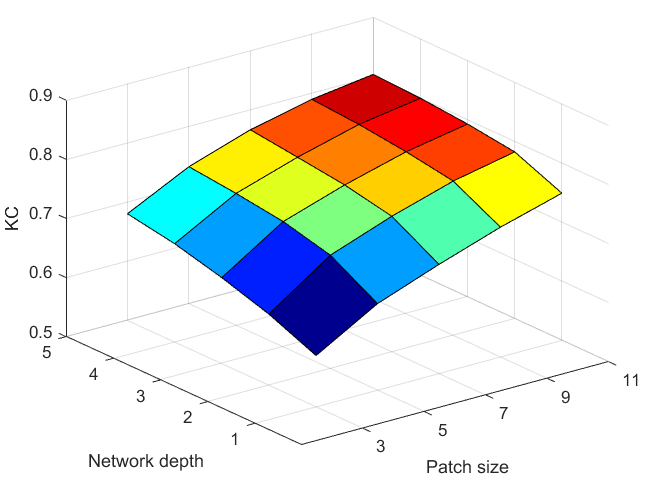}
  \label{fig_QU}}

  \caption{The performance of the KPCA-MNet with different network depths $L$ and convolution kernel sizes $w$. (a) WH data set. (b) QU data set.}
  \label{fig:dis_LW}
\end{figure}

\par Compared with all the methods mentioned before, SVM achieves a better binary change map, even though a few pixels are misclassified. The result obtained by DSCN is similar to that obtained by PCA-Kmeans, yet more changed regions are falsely detected as unchanged ones. In Fig. 7-(j), these main changed regions are correctly classified by RNN-CD. But some details in changed regions are missed and plenty of unchanged buildings and roads are falsely detected. In the result of SAE, a majority of changed pixels are correctly detected. Nonetheless, a part of unchanged features is incorrectly classified as change, such as vegetation. Finally, there is undeniable that the proposed KPCA-MNet generates the best qualitative result with less noise and more complete changed regions. 

\par Table III reports the values of evaluation criteria on the QU data set. Compared with CVA and PCA-Kmeans, MAD and SFA perform better in the QU data set. Through adopting objects as the basic units of CD to utilize spatial context information, the FP of OBCD is the best again. But due to only using the low-level features, numerous changed pixels are incorrectly detected as unchanged ones, which makes the OE, OA, and KP of OBCD unsatisfactory. Finally, it can be observed that KPCA-MNet outperforms the other compared methods with FN of 1795, OE of 3468, OA of 0.9019, and KC of 0.7778, which proves the effectiveness of KPCANet for binary CD in VHR images once again. 

\par Furthermore, it is worth noting that, in the WH data set, trained in the samples selected by the unsupervised pre-detection method, RNN-CD achieves a relatively good result. Nonetheless, in the QU data set, the performance of RNN-CD is even worse than MAD and SFA. This is because the automatic selection of training samples is unsupervised and some false alarm samples may be included in the training phase. Therefore, if the automatic pre-detection cannot generate relatively clean and representative training samples in a data set, it is difficult for the deep learning method based on pre-detection to show good performance in this data set. Besides, the accuracy of DSCN on both data sets is not high, which implies that this pre-detection-based approach does not apply to all deep learning models. In contrast, the training phase of KPCA-MNet does not require annotated data and the performance of KPCA-MNet is better and more stable. 

\begin{figure}[t]

  \centering
  \includegraphics[scale=0.5]{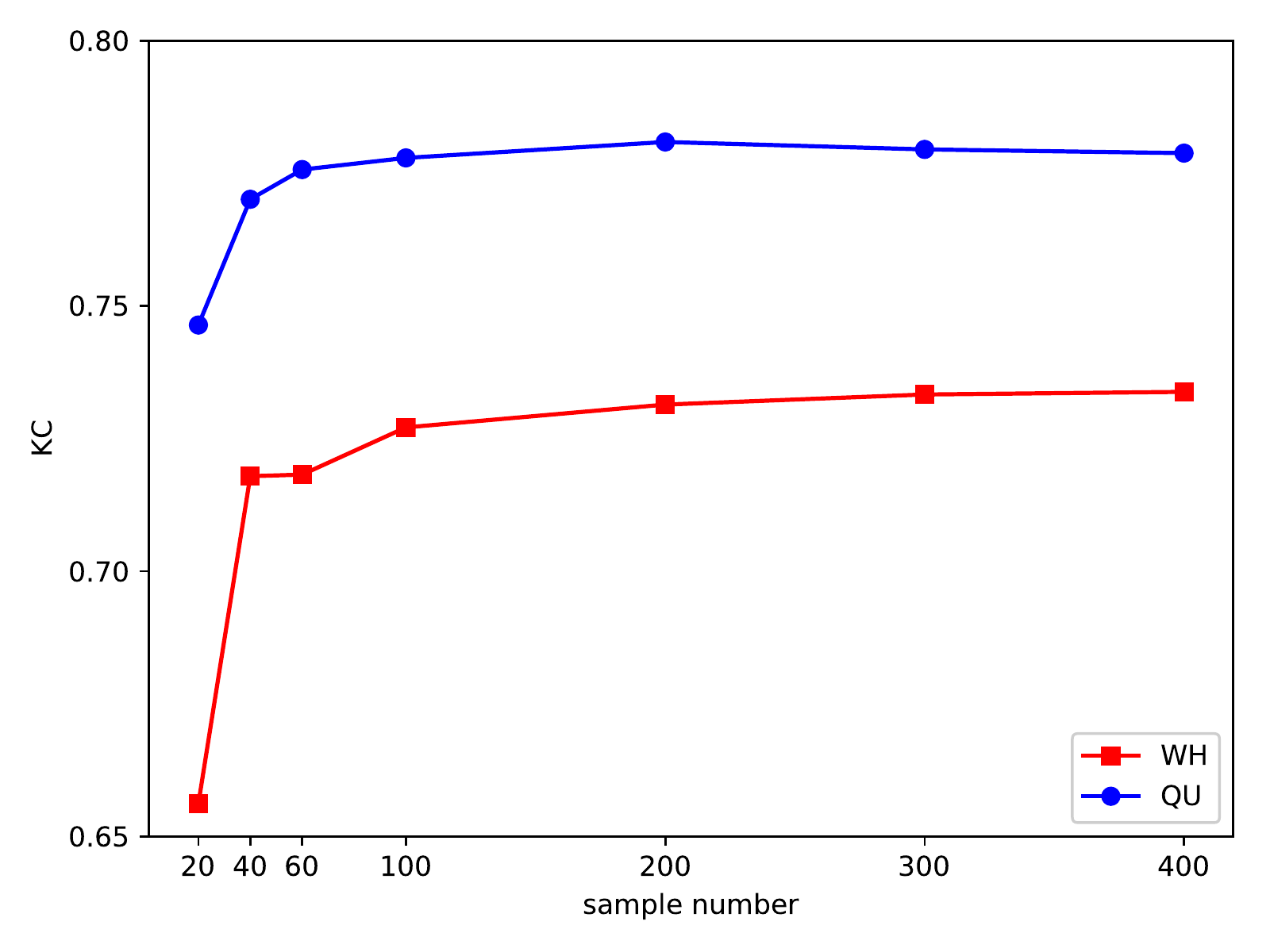}
  \caption{The performance of the KPCA-MNet with different number of training samples $n$.}
  \label{fig:dis_sample_num}
\end{figure}

\subsection{Discussion}
\par In both data sets, the KPCA-MNet achieves good performance and obtains accurate binary change maps. And there are several parameters of KPCA-MNet play an important role, including the selection of kernel function $k\left(x, y\right)$, the number of training patches $n$, the number and size of KPCA convolution kernel, $p$ and $w$, and the network depth $L$. For the sake of simplify, we employ the same parameters per layer to reduce the vast number of possible choices.

\par The network depth $L$ and the convolution kernel size $w$ are first discussed. As shown in Fig. \ref{fig:dis_LW}, the deeper network shows a better performance in both data set. But considering that KPCA-MNet is trained in a layer-wise manner, a too deep architecture (deeper than 4) would increase the burden of computing and not help to increase accuracy too much. Therefore, for network depth $L$, 2 to 4 would be good choices. 

\par For the parameter $w$, the larger w will increase the receptive field of the network and makes the network utilize more spatial context information. However, if $w$ is too large, some unrelated information is considered by the network, which would damage CD performance. Therefore, in the WH data set with a spatial resolution of 4m, the KP increases greatly as the $w$ increases from 3 to 5. But when $w$ is larger than 5, the accuracy decreases gradually as the $w$ gets larger. For the QU data set, the spatial resolution of images is 2.4m, the performance of the network becomes better as the $w$ gets larger. However, when $w$ greater than 9, the network performance improvement becomes no longer obvious. Specifically, depending on the spatial resolution of different VHR images, 20m to 30m is a suitable range for the real size of the KPCA convolution kernel. 

\begin{figure}[t]

  \centering
  \includegraphics[scale=0.5]{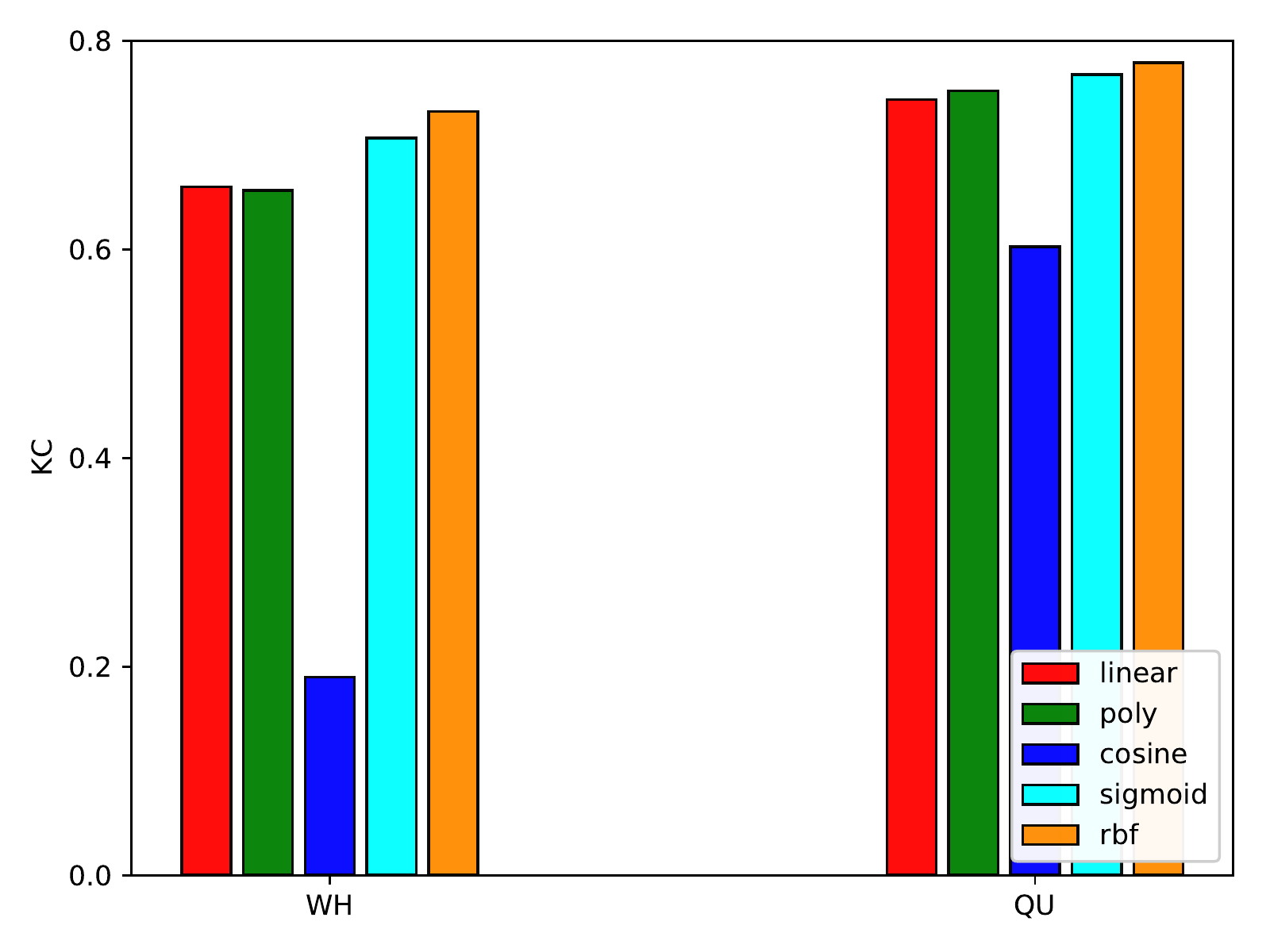}
  \caption{The performance of the KPCA-MNet with different kernel function $k\left(x, y\right)$.}
  \label{fig:dis_kernel_acc}
\end{figure}

\par The number of training samples $n$ will affect the feature extraction ability of KPCA convolution. The more training samples are, the spatial-spectral features extracted by the KPCA convolution are more representative. However, if $n$ is too large, the size of the constructed kernel matrix is too large, which will increase computational cost and even cause the KPCA problem to be difficult to solve. As shown in Fig. \ref{fig:dis_sample_num}, setting $n = 200$ is a good choice.

\par For the KPCA convolution, the selection of kernel function is significant, because the different kernel functions have diverse mapping ability. In Fig. \ref{fig:dis_kernel_acc}, the performance of the KPCA-MNet with five commonly used kernel function in both data sets is compared. RBF is the best choice for KPCA convolution, which can make the network extract more representative nonlinear features. And the sigmoid kernel also shows good performance. Compared with linear PCA convolution, selecting RBF or sigmoid as a kernel function can significantly improve the CD performance, which proves the necessity of generalizing linear PCA convolution to KPCA convolution. 

\par In addition, the number of KPCA convolution kernel $p$ is also related to the CD performance. The more convolution kernels can extract more features from the multi-temporal images for CD. As a trade-off between accuracy and computational expense, we empirically set parameter $p$ as 8.

\section{Experiment of Multi-class Change Detection}\label{sec:5}

\subsection{Data Description}
\par
\begin{figure}[t]
  \centering

  \subfloat[]{
    \includegraphics[width=1.05in]{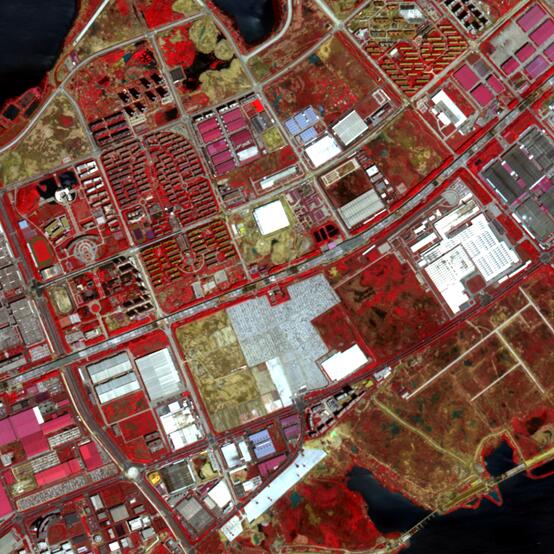}
  \label{fig_first_case}}
  \hfil
  \subfloat[]{
    \includegraphics[width=1.05in]{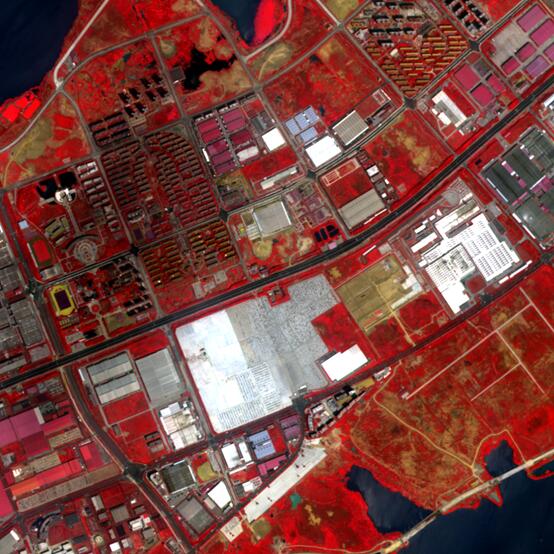}
  \label{fig_second_case}}
  \hfil
  \subfloat[]{
    \includegraphics[width=1.05in]{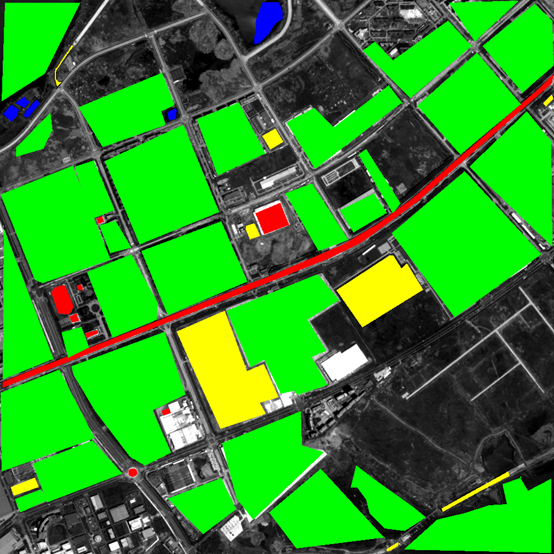}
  \label{fig_third_case}}

  \caption{The pseudo-color VHR images of (a) HY-1, (b) HY-2. (c) is ground truth, where red indicates city change, blue indicates water change, yellow indicates soil change and green indicates unchanged region.}
  \label{HY_dataset}
\end{figure}

 Not only can be used for binary CD, but the KPCA-MNet also can detect multi-class change between multi-temporal VHR images. In the experiment of multi-class CD, HY data set is adopted to evaluate the proposed method. The HY data set was acquired by GF-2 sensor over the city of Hanyang, China, in April 2016 and September 2016. These two images both consist of 1000 1000 pixels with a spatial resolution of 4m. Owing to the rapid development of Hanyang, the study area shows obvious land-cover changes and the changes between the two images mainly involve city change (buildings or roads to other artificial facilities), water change (water regions to bare soils or vegetation), and soil change (soils to artificial facilities or water regions). The pseudo-color VHR images and the corresponding ground truth of three multi-class change types and non-changes are shown in Fig. \ref{HY_dataset}. In the ground truth, red indicates city change, blue indicates water change, yellow indicates soil change, green indicates unchanged area, and the remaining pixels are undefined. What’s more, the HY data set is also suffered from the “over-exposed” problem. 

\subsection{Experiment Setting}
\par According to the conclusion derived from section IV-D, in the experiment of multi-class CD, RBF is chosen as the kernel function, the number of training patches $n$ is set as 200, the size of KPCA convolution kernel $w$ is set as 5, setting parameter $p$ as 8, and two weight-shared KPCA convolutional layers are stacked to extract spatial-spectral features. 

\par For the purpose of demonstrating the superiority of the KPCA-MNet in unsupervised multi-class CD, seven representative methods are adopted for comparison: PCC, 2D-CVA \cite{Sharma2007}, $C^{2}$VA \cite{Bovolo2012}, PCA, IRMAD \cite{Nielsen1997}, SAE \cite{Bengio2007}, and RandNet. Specifically, PCC is a widely used framework for multi-class CD. In the experiment, the unsupervised clustering algorithm, ISODATA, is utilized to get the multi-temporal classification maps. In 2D-CVA, two bands are selected to calculate the magnitudes and directions of spectral change vectors to detect diverse kinds of change. The optimal combination can be selected by trail-and-test or some priori information. $C^{2}$VA compresses all the bands into the magnitudes and directions of spectral change vectors and perform threshold segmentation and clustering to get the multi-class CD result. In PCA-based approach, the magnitude and directions of the first two principal components are computed to generate the multi-class change map. As for IRMAD, it is similar to 2D-CVA and PCA, except that the last two MAD variates are used to calculate the magnitude and direction since they contain the most change information. In SAE, spatial-spectral features are first extracted from VHR images, then the same way proposed in $C^{2}$VA is adopted to utilize these features for CD. To evaluate the effectiveness of the proposed model, a natural idea is to simplify the KPCA convolution kernels in KPCA-MNet by random convolution kernels following the standard Gaussian distribution, thus a variant of KPCA-MNet, RandNet is introduced, which has the same structure as KPCA-MNet but replaces the KPCA convolution kernels with the random filters following the standard Gaussian distribution. What’s more, to prove the necessity of generalizing linear PCA convolution to KPCA convolution, a variant of the proposed method entitled LinearPCA-MNet is developed by replacing KPCA convolution layers with linear PCA convolution layers.

\par For the accuracy assessment, we first measure the accuracy of per-class, and then OA and KP are used to measure the overall performance. 

\subsection{Experimental Result and Analysis}
\par
\begin{figure*}[t]
  \centering
  \subfloat[]{
    \includegraphics[width=1.9in]{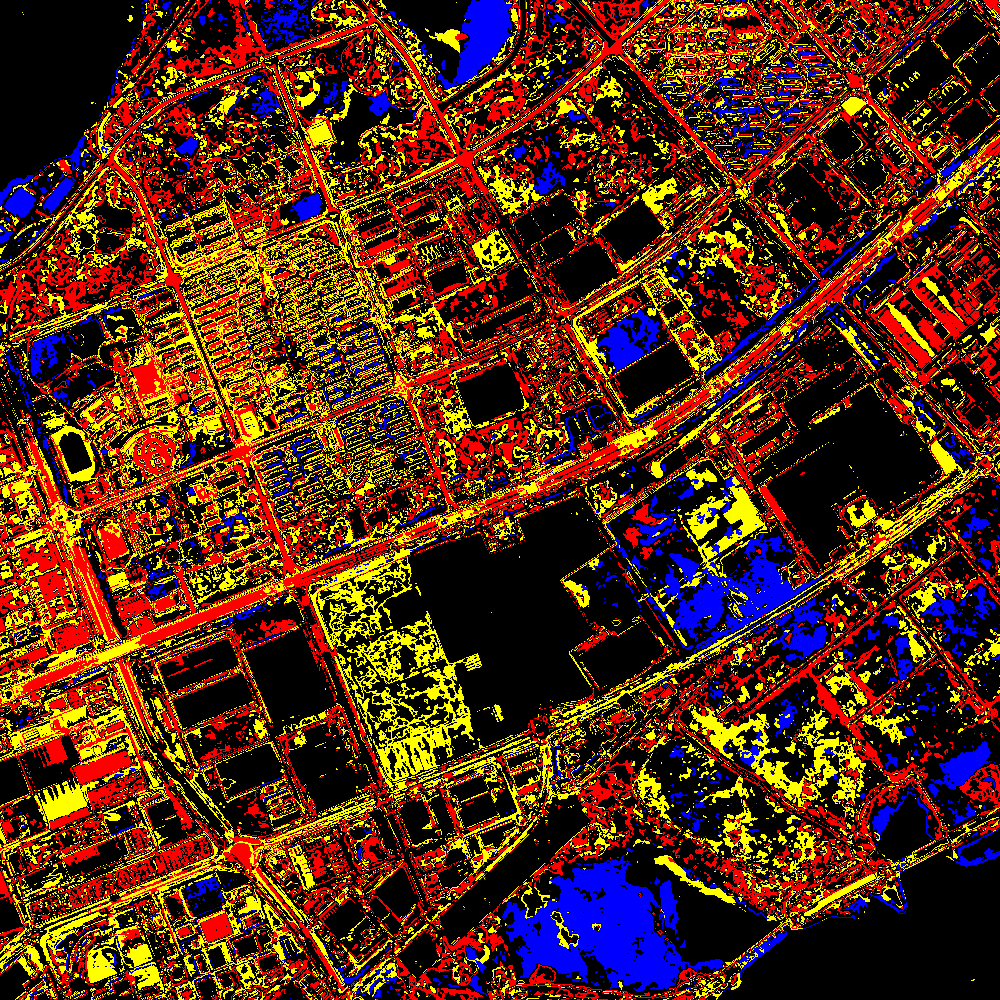}
  \label{HY_PCC}}
  \hfil
  \subfloat[]{
    \includegraphics[width=1.9in]{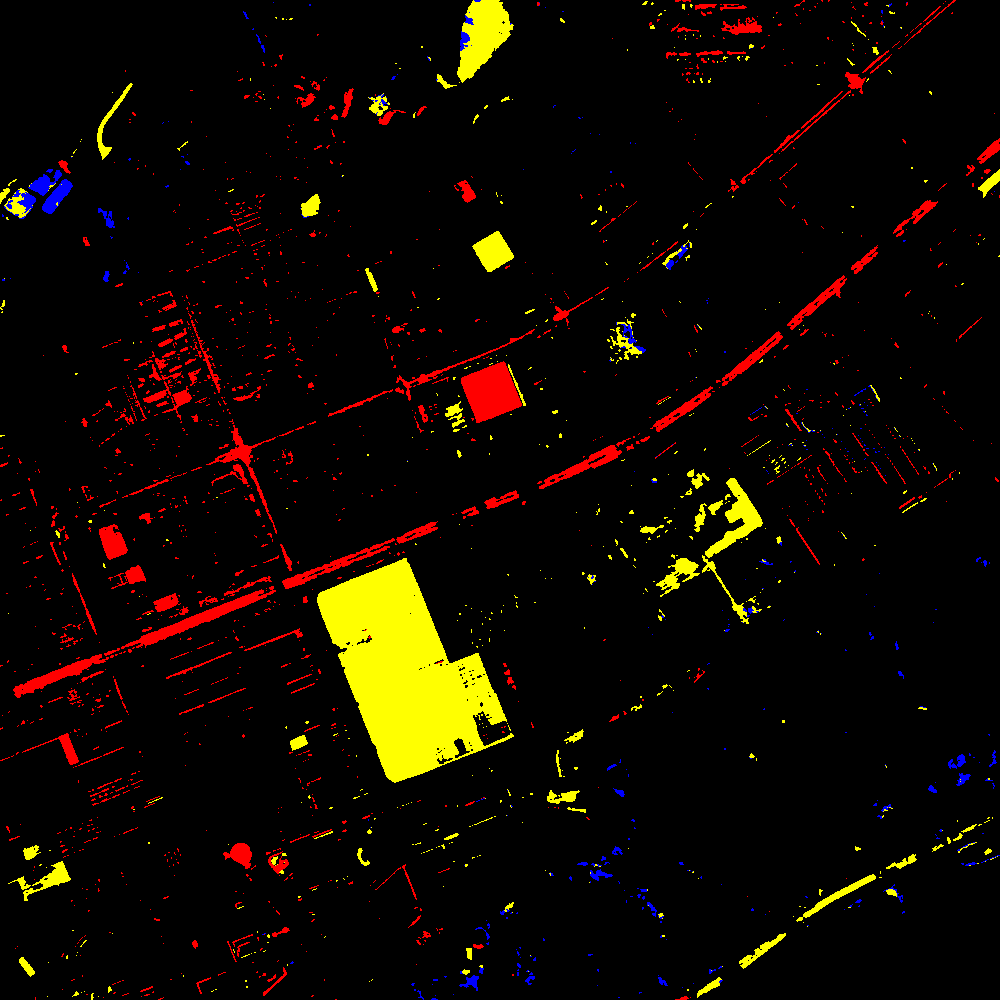}
  \label{HY_CVA-2D}}
  \hfil
  \subfloat[]{
    \includegraphics[width=1.9in]{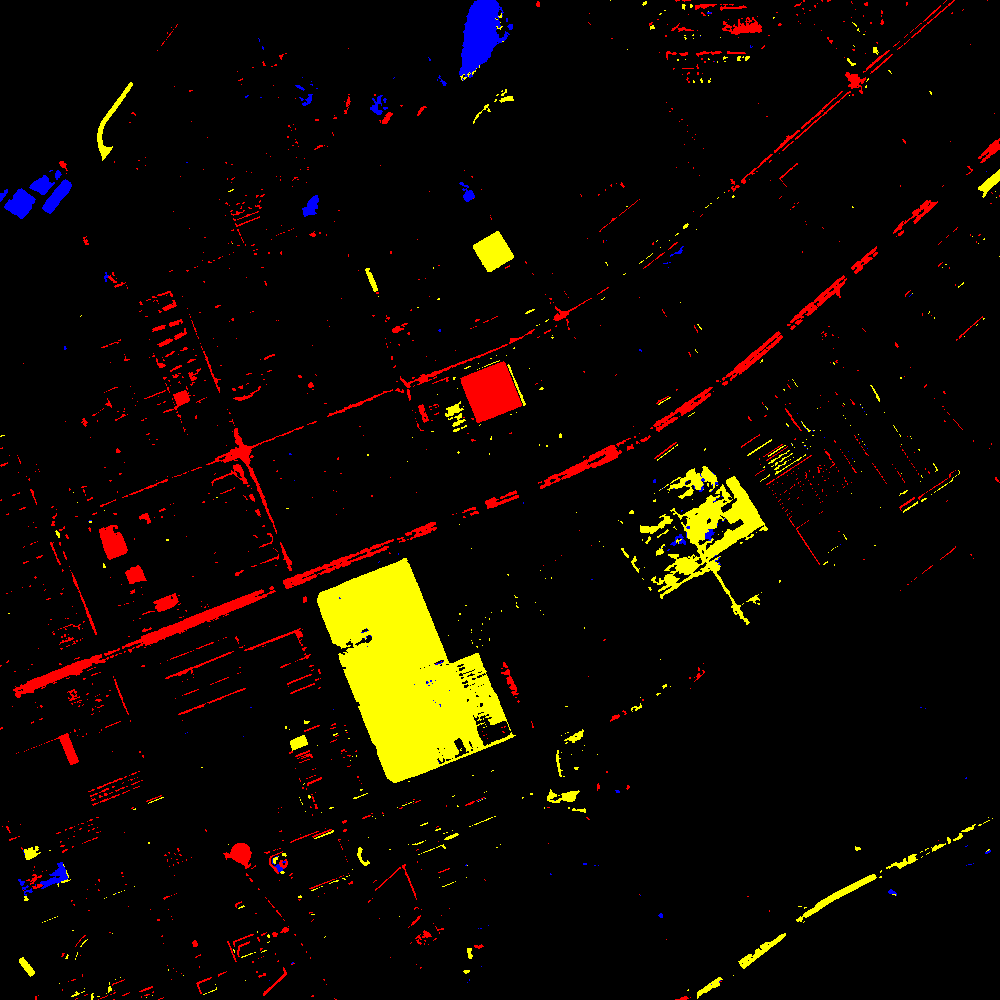}
  \label{HY_C2VA}}
  
  \subfloat[]{
    \includegraphics[width=1.9in]{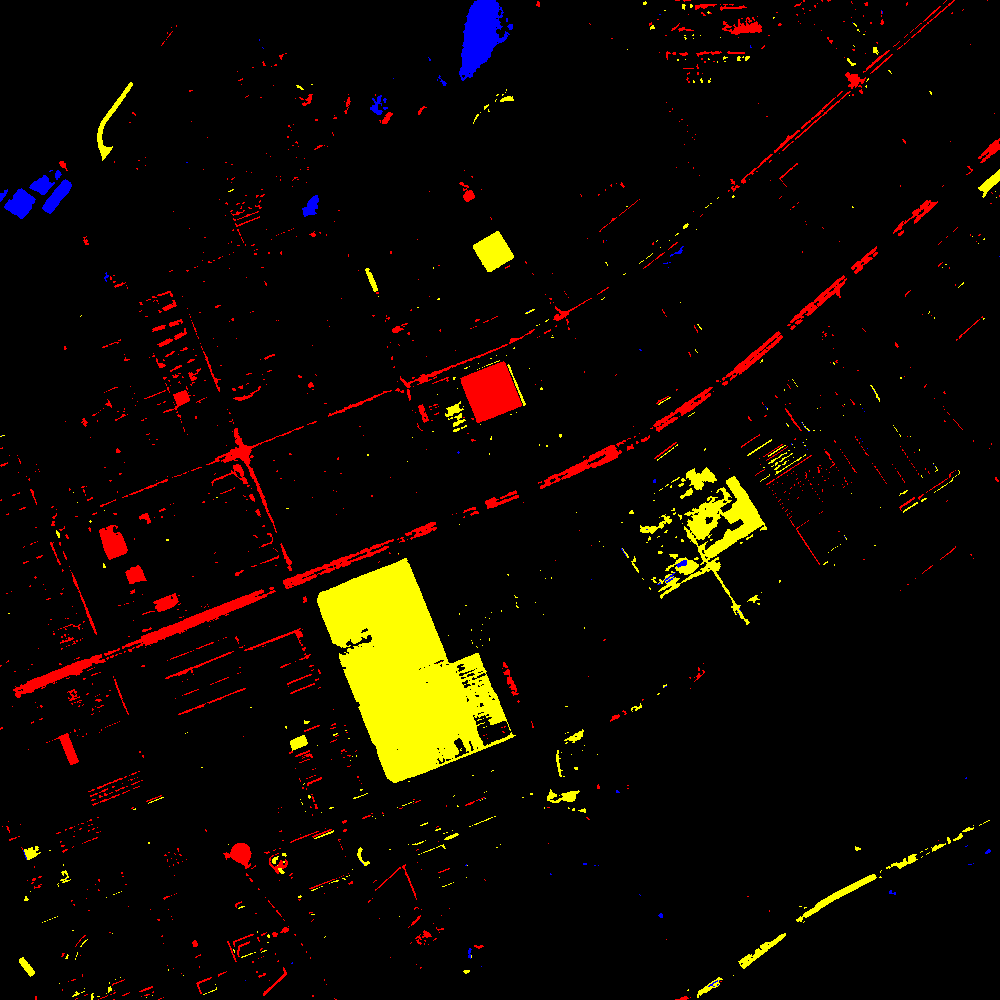}
  \label{HY_PCA}}
  \hfil
  \subfloat[]{
    \includegraphics[width=1.9in]{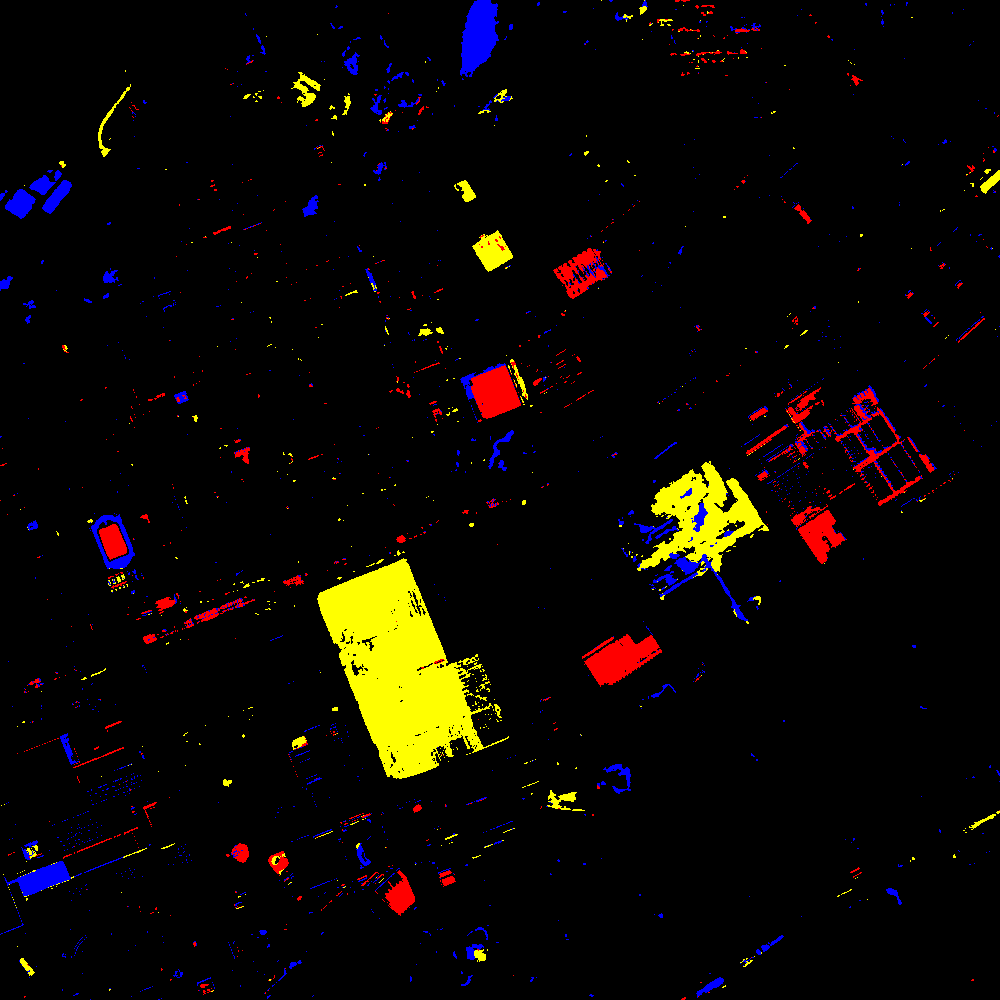}
  \label{HY_IRMAD}}
  \hfil
  \subfloat[]{
    \includegraphics[width=1.9in]{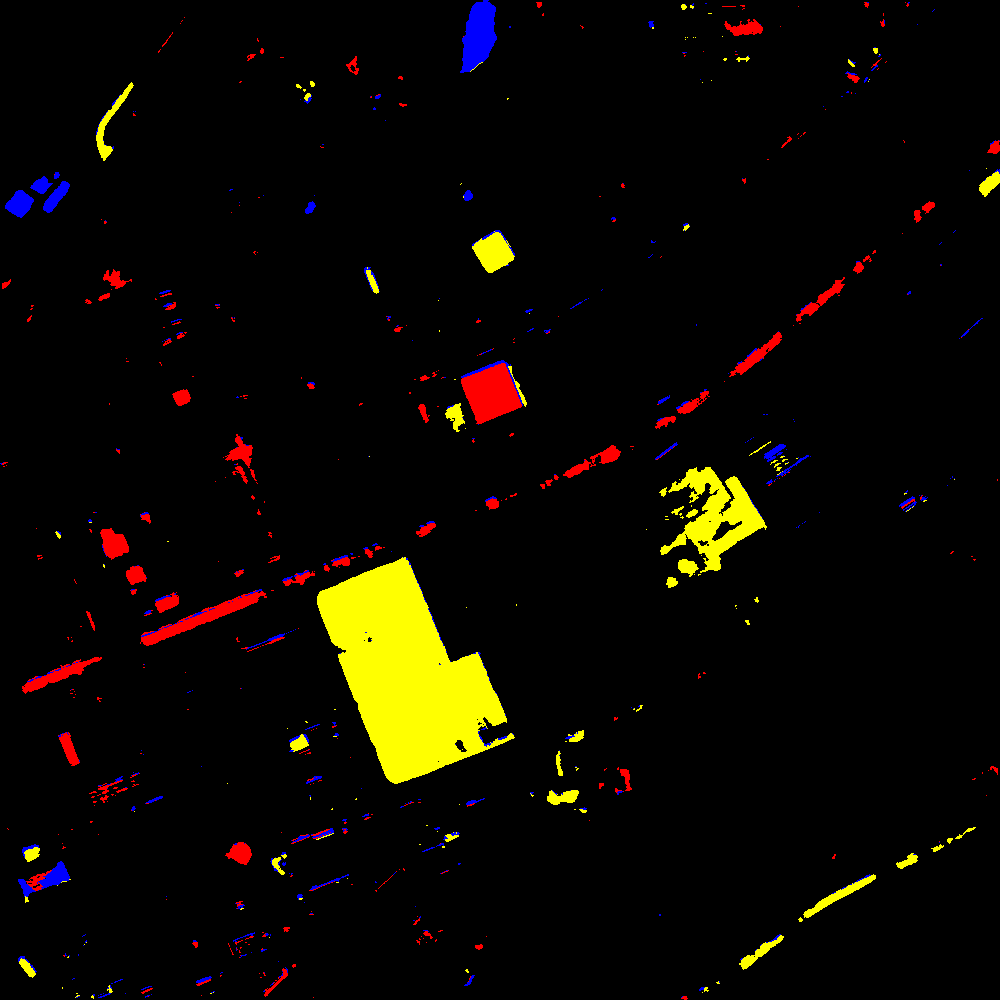}
  \label{HY_SAE}}

  \subfloat[]{
    \includegraphics[width=1.9in]{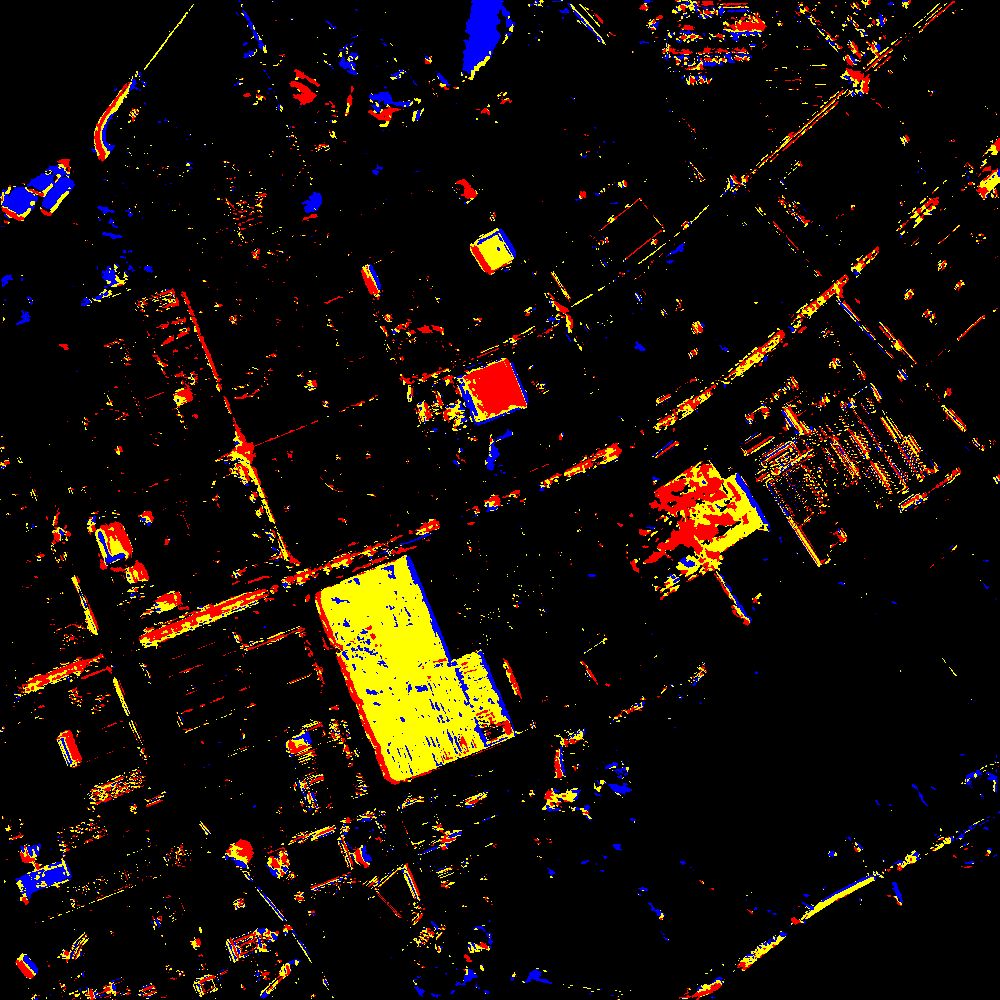}
  \label{HY_RandNet}}
  \hfil
  \subfloat[]{
    \includegraphics[width=1.9in]{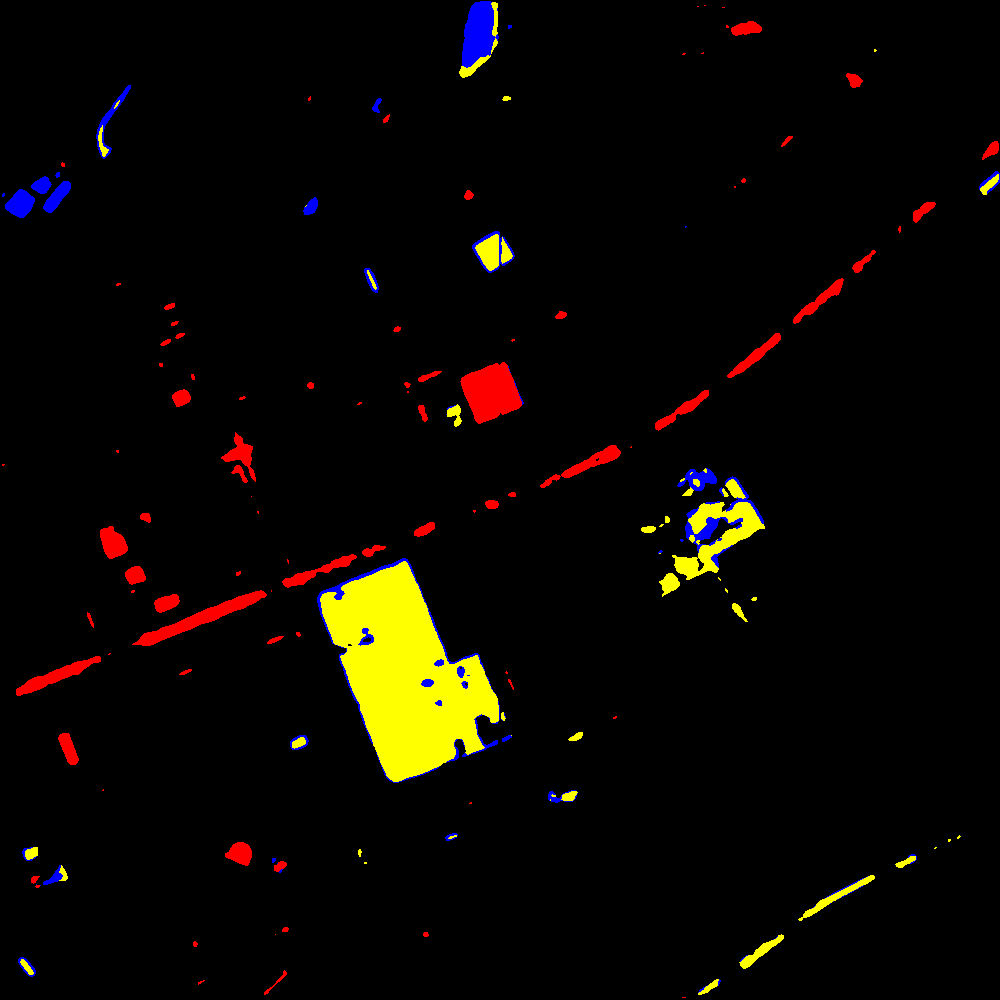}
  \label{HY_Linear-KPCANet}}
  \hfil
  \subfloat[]{
    \includegraphics[width=1.9in]{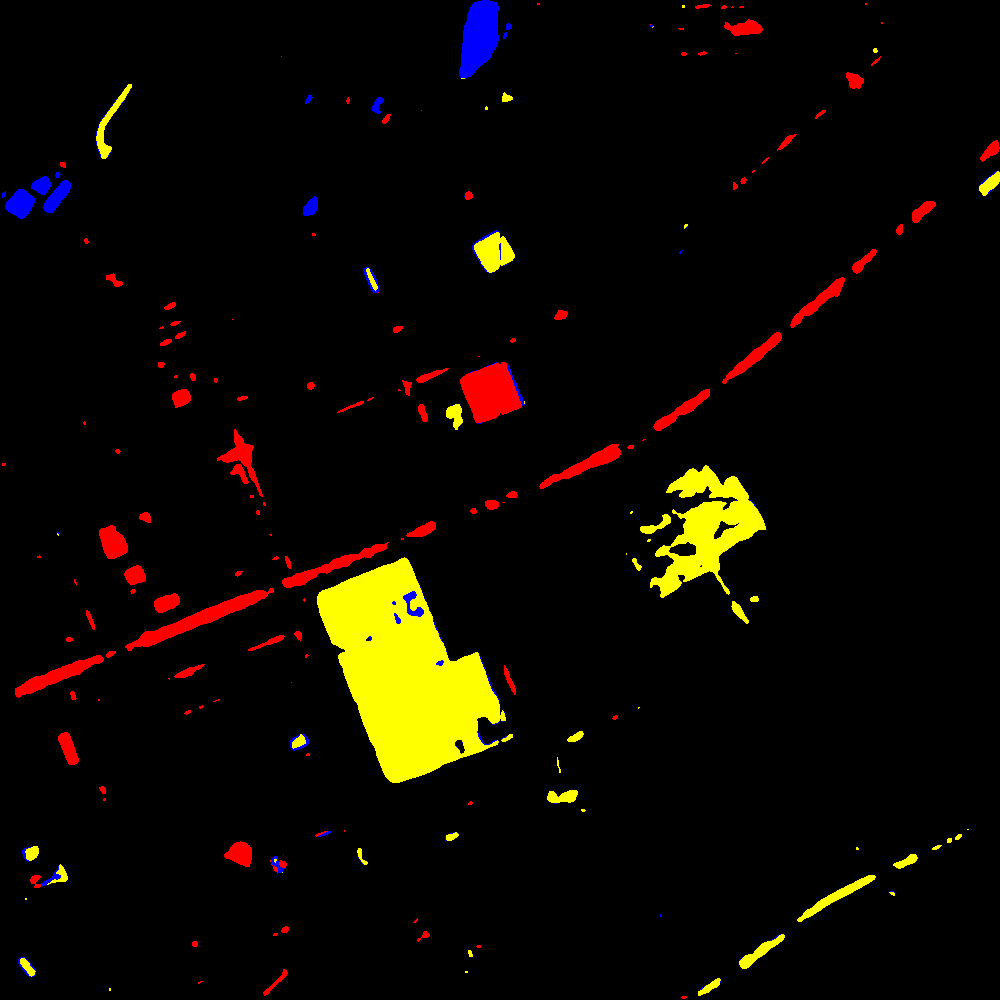}
  \label{HY_KPCAMNet}}
  
  \caption{Multi-class change maps obtained by the proposed method and comparison methods on the HY data set. (a) PCC. (b) 2D-CVA. (c) C 2VA. (d) PCA. (e) IRMAD. (f) SAE. (g) RandNet. (h) LinearPCA-MNet. (i) KPCA-MNet.}
  \label{HY_result}
\end{figure*}

 Fig. \ref{HY_result} shows the multi-class CD results acquired by the proposed method and comparison methods on the HY data set. The change map generated by PCC is not good in visual, because the clustering techniques are powerless in classify the land-cover accurately, which would accumulate the false detection to result in a rough change map. In the results of 2D-CVA, the major changed and unchanged regions are correctly detected. However, due to the changed information from only two bands is used, water changes are unclassified as soil changes and some soil changes are not detected. By the magnitude and direction of spectral change vector, change information from all bands is compressed into a polar representation. Therefore, the multi-class change map obtained by $C^{2}$VA is better than the ones obtained by 2D-CVA. But due to some information is lost after the compress procedure, a part of soil changes is falsely detected as water changes. Only spectral information is utilized by $C^{2}$VA, thus all kinds of changed regions have internal fragmentation and unchanged regions have salt-and-pepper noise. Besides, caused by the “over-exposed” problem, a few unchanged regions are incorrectly classified as soil changes and city changes. For PCA, because the main change information is concentrated in the first two principal components, the CD result of PCA is similar to the ones of $C^{2}$VA. In Fig. 12-(e), the main soil changes and water changes are correctly classified by IRMAD, but most of the city changes are not detected. What’s more, IRMAD is strongly puzzled by the “over-exposed” problem. 

\par Compared with $C^{2}$VA, the spatial-context information is utilized to detect multi-class changes by SAE, thus, most of the unchanged regions are classified correctly and there are a little internal fragmentation and salt-and-pepper noise in the result. Nonetheless, the learning process of SAE can only make the essential features of VHR images be extracted, it cannot guarantee that the extracted features are sufficiently discriminative for distinguishing different kinds of changes. Therefore, some soil changes and plenty of city changes are not detected by SAE. As shown in Fig. 12-(g), though sharing the same architecture with KPCA-MNet, RandNet performs not well in the HY data set. A part of unchanged regions is misclassified as change and the three kinds of changes are mixed together, which indicates that the random convolution kernels cannot extract valid features for multi-class CD.

\par The multi-class change map detected by the LinearPCA-MNet is good. However, the features extracted by linear PCA convolution are not representative enough, which results in the misclassification between soli changes and water changes. By contrast, in the result of KPCA-MNet, most of the changed regions and unchanged regions are detected accurately and three kinds of changes are efficiently distinguished from each other, which implies that the KPCA convolutions do extract representative spatial-spectral features from VHR images and the different kinds of changes are efficiently discriminative in the new feature space represented by FM and WFD. The visual comparisons between multi-class change maps fully demonstrate the effectiveness of the proposed KPCA-MNet.

\par Table \ref{HY_table} lists the quantitative statistics on the change maps by different methods. And the confusion matrices of CD results are shown in Fig. 12. Except for the water changes accuracy, each metric of PCC is not high. The accuracy achieved by IRMAD and RandNet is also not good, the KPs of these two methods are 0.6360 and 0.6062, respectively. 2D-CVA achieves a relative high city changes accuracy, but a major part of water changes is not classified incorrectly, which leads to low accuracy of water change with 0.2551. Since utilizing the changed information in each band, the performance of $C^{2}$VA is better than 2D-CVA with a higher KP of 0.7634. SAE achieves high accuracy, with OA of 0.9576 and KP of 0.7712. But the accuracy of city changes in SAE is not good with only 0.4797. Compared with these methods, the KPCA-MNet behaves prominent superiority in all evaluation criteria, which demonstrate that the proposed method is applicable to unsupervised multi-class CD in multi-temporal VHR images. What’more, though the LinearPCA-MNet shows good performance in multi-class CD with KP of 0.7745, it still cannot compete with the KPCA-MNet, especially in the accuracy of each change class, which proves that the KPCA convolution has more powerful feature extraction ability compared to the linear PCA convolution.

\begin{table*}[t]
  \renewcommand{\arraystretch}{1.4}
  \caption{Accuracy assessment on multi-class change maps obtained by different methods on the HY data set}
  \label{HY_table}
  \centering
  \begin{tabular}{c c c c c c c}
    \hline
    \multirow{2}{*}{Method} & \multicolumn{4}{c}{Per-class} & \multirow{2}{*}{OA} & \multirow{2}{*}{KC} \\
    \cline{2-5}
     &  soil &  water &  city & non-change & & \\
    \hline\hline
    PCC	& 0.3530	& 	\textbf{0.9052}	& 0.5043 & 0.6633	& 0.6375	& 0.1313 \\ 	
    2D-CVA	& 0.7138	& 	0.2551	& \underline{0.6330} & 0.9796	& 0.9428	& 0.7089 \\ 									
    $C^{2}$VA	& 0.7546	& 	0.8784	& 0.6027 & 0.9841	& 0.9538	& 0.7634 \\ 															
    PCA	& 0.7320	& 	\underline{0.8926}	& 0.5990 & 0.9848	& 0.9528	& 0.7546 \\ 	 							
    IRMAD & 0.7200	& 	0.8389	& 0.2230 & 0.9744	& 0.9294	& 0.6360 \\ 	
    SAE	& \underline{0.7646}	& 	0.7443	& 0.4797 & 0.9936	& 0.9576	& 0.7712 \\ 						
    RandNet	& 0.6556	& 	0.7825	& 0.3834 & 0.9544	& 0.9120	& 0.6062 \\ 		
    LinearPCA-MNet & 0.6890	& 0.7393	& 0.5689	& \textbf{0.9975}	& \underline{0.9586} & \underline{0.7745} \\ 	
    KPCA-MNet	& \textbf{0.8008}	& 0.8862	& \textbf{0.6438}	& \underline{0.9952}	& \textbf{0.9687} & \textbf{0.8334} \\ 			
    \hline
  \end{tabular}
\end{table*}

\subsection{Discussion}
\par
\begin{figure}[t]
  \centering

  \subfloat[]{
    \includegraphics[width=1.5in]{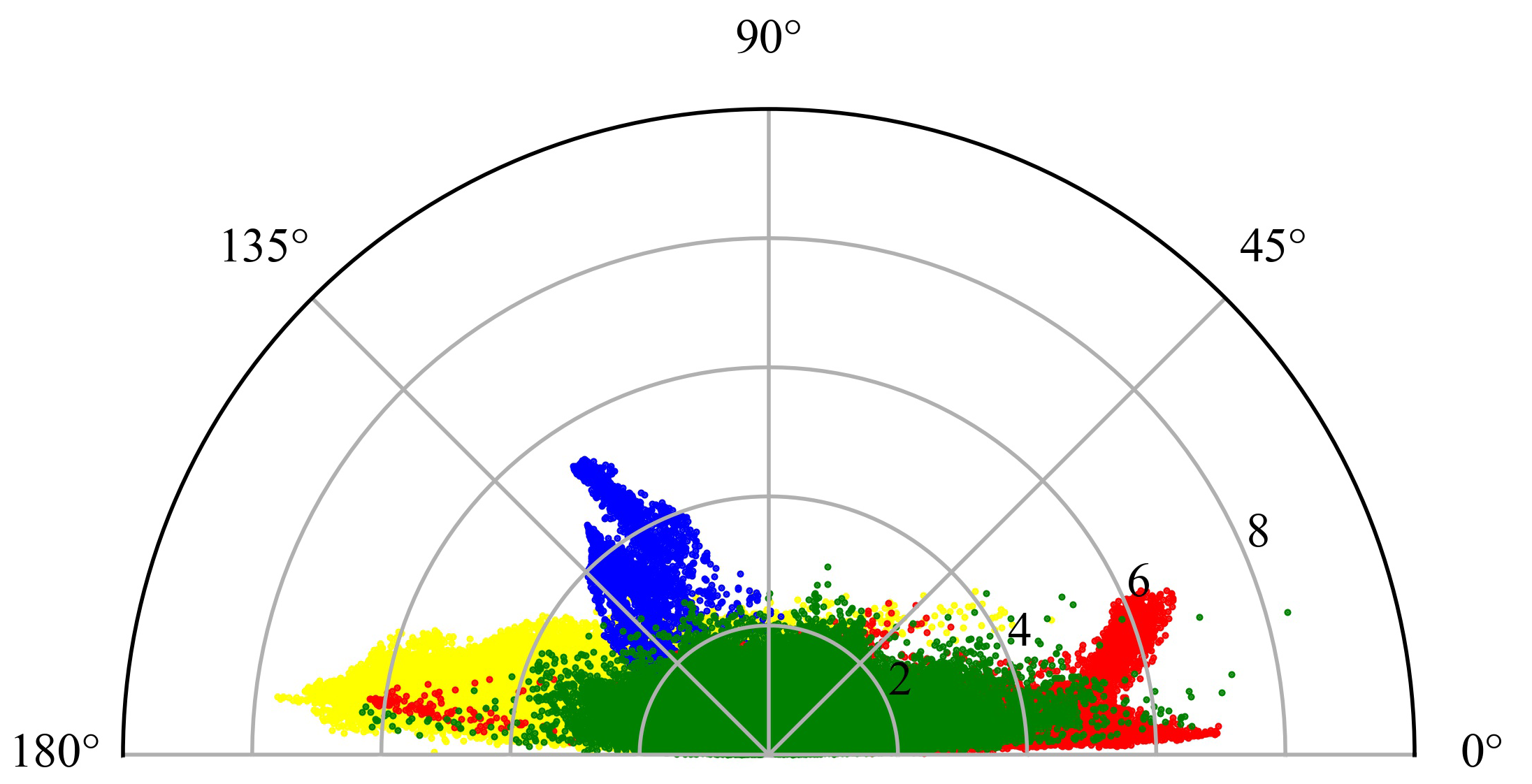}
  \label{fig_first_case}}
  \hfil
  \subfloat[]{
    \includegraphics[width=1.5in]{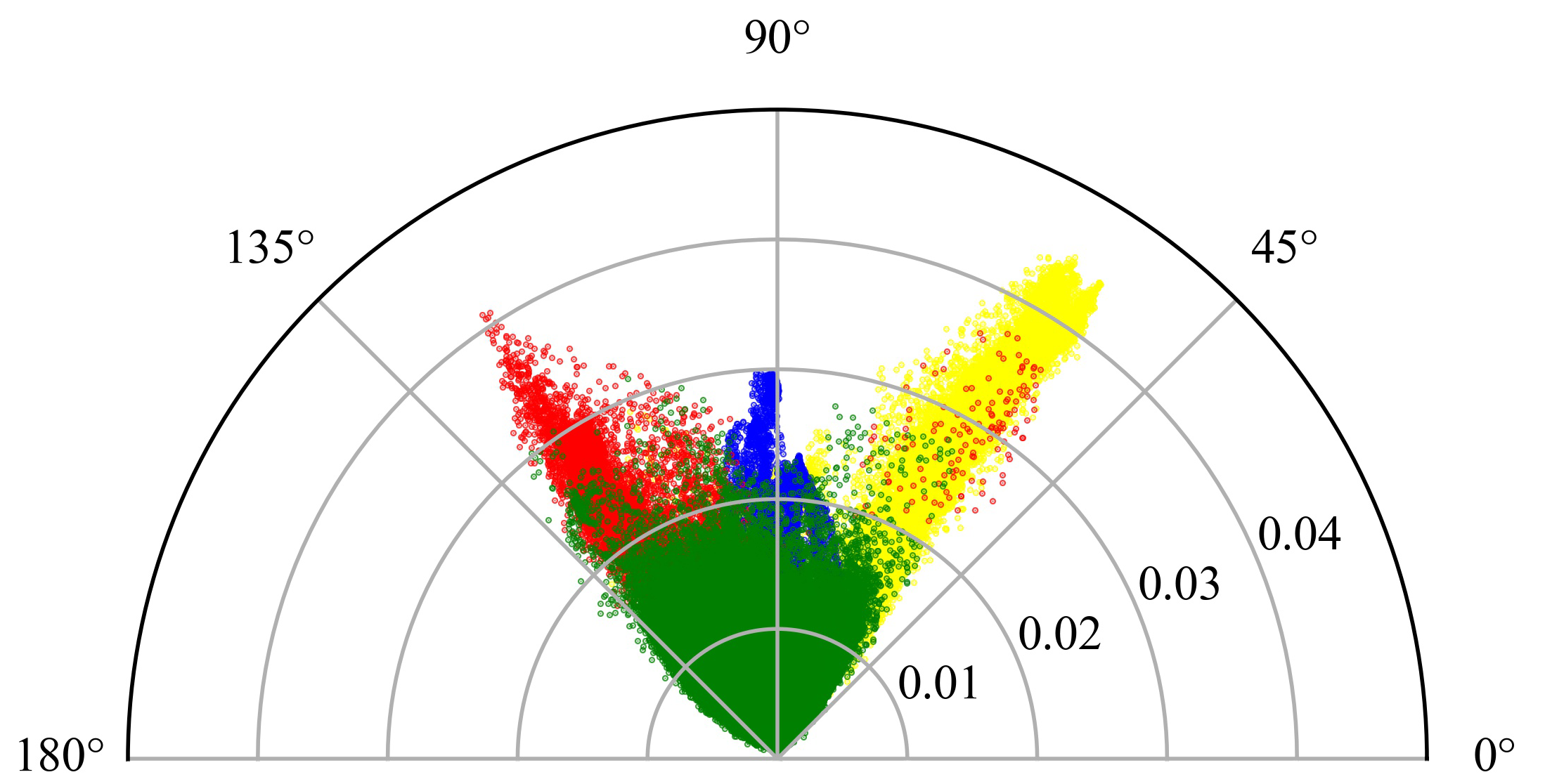}
  \label{fig_second_case}}
  
  \subfloat[]{
    \includegraphics[width=1.5in]{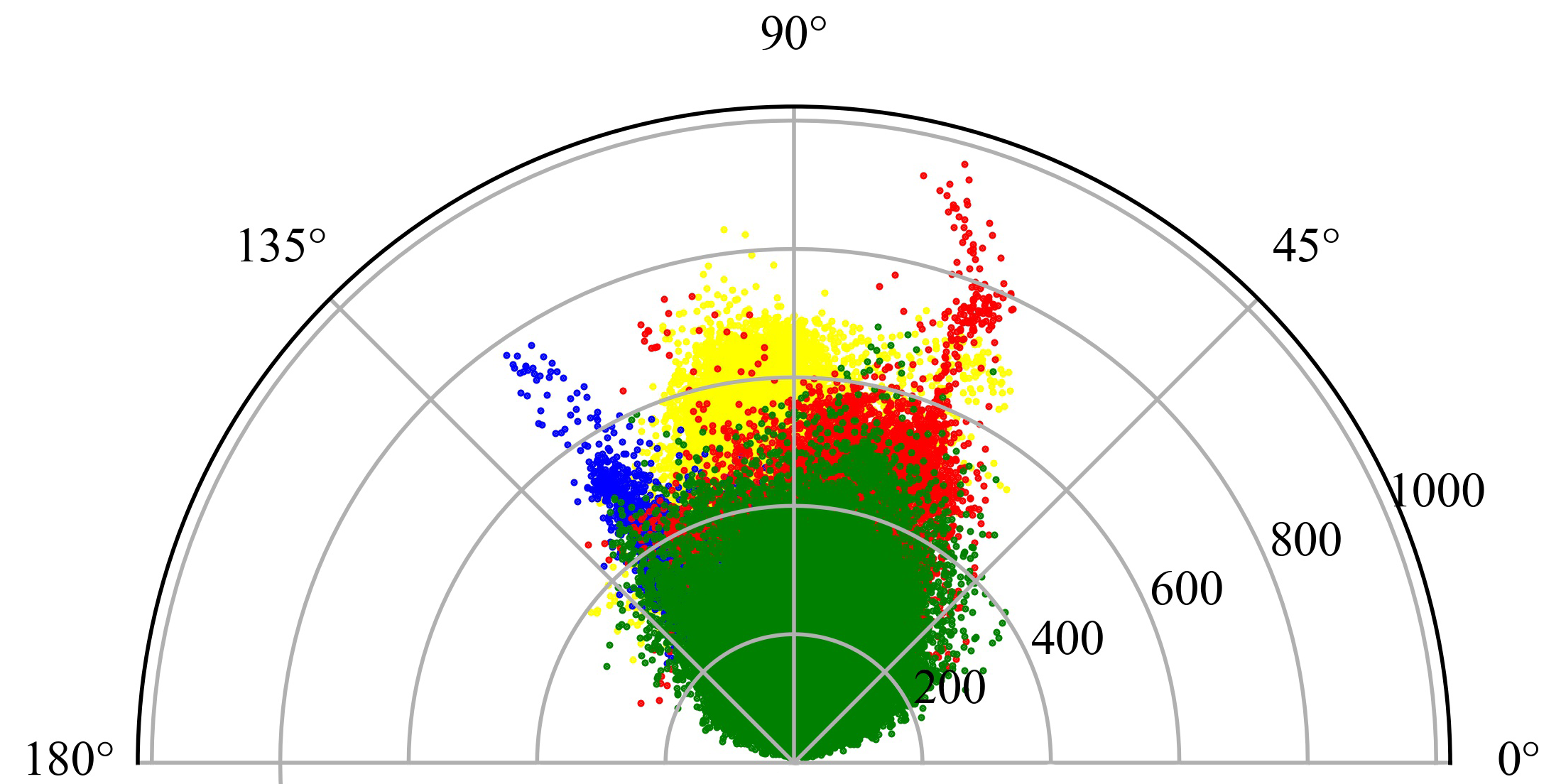}
  \label{fig_third_case}}
  \hfil
  \subfloat[]{
    \includegraphics[width=1.5in]{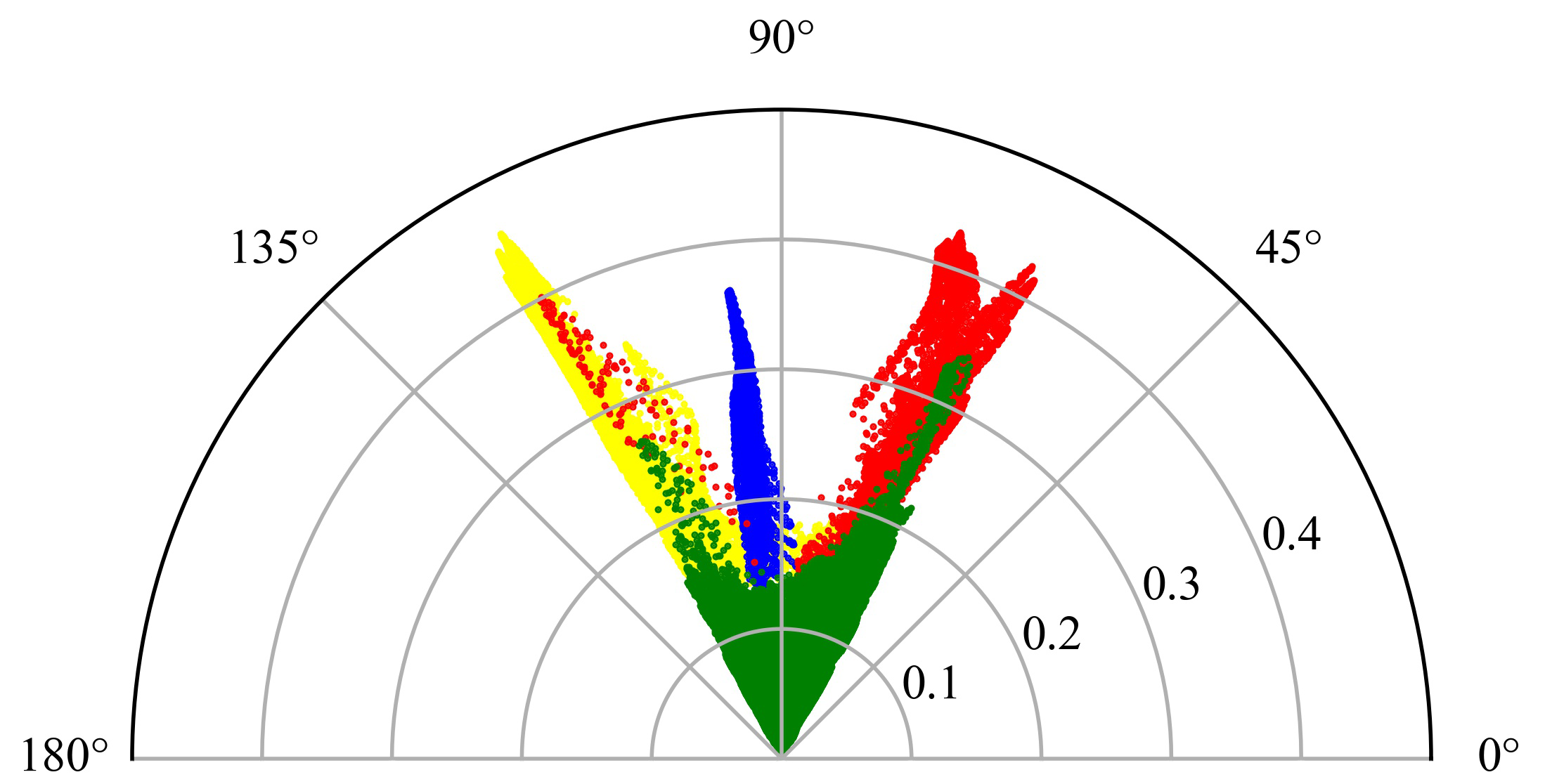}
  \label{fig_fourth_case}}

  \caption{Distribution of the reference data in the 2-D polar domain of different methods, where red indicates city change, blue indicates water change, yellow indicates soil change and green indicates non-change. (a) $C^{2}$VA. (b) SAE. (c) RandNet. (d) KPCA-MNet.}
  \label{dis_2dpolar}
\end{figure}

\begin{figure*}[t]
  \centering

  \subfloat[]{
    \includegraphics[width=1.6in]{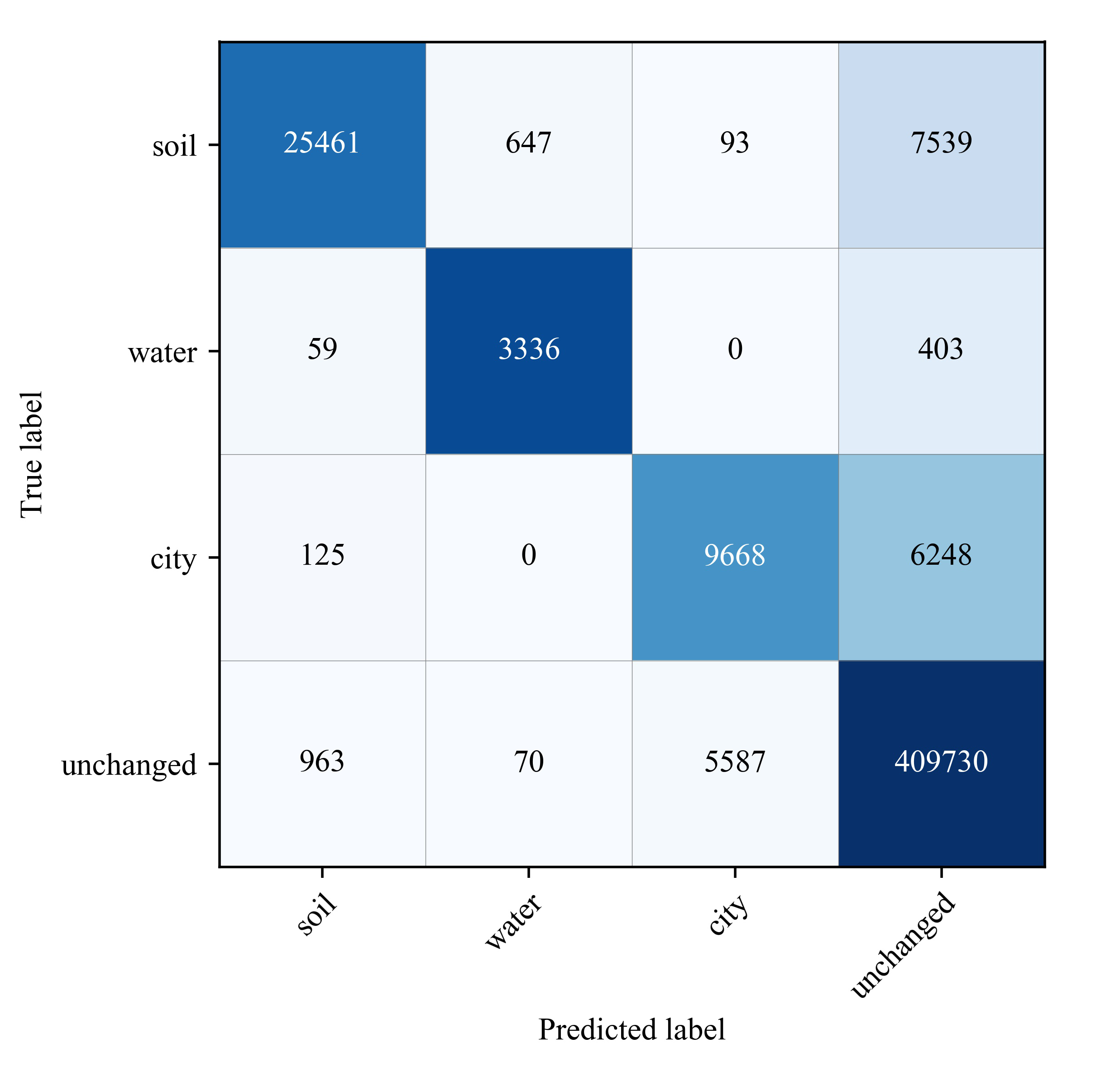}
  \label{fig_first_case}}
  \hfil
  \subfloat[]{
    \includegraphics[width=1.6in]{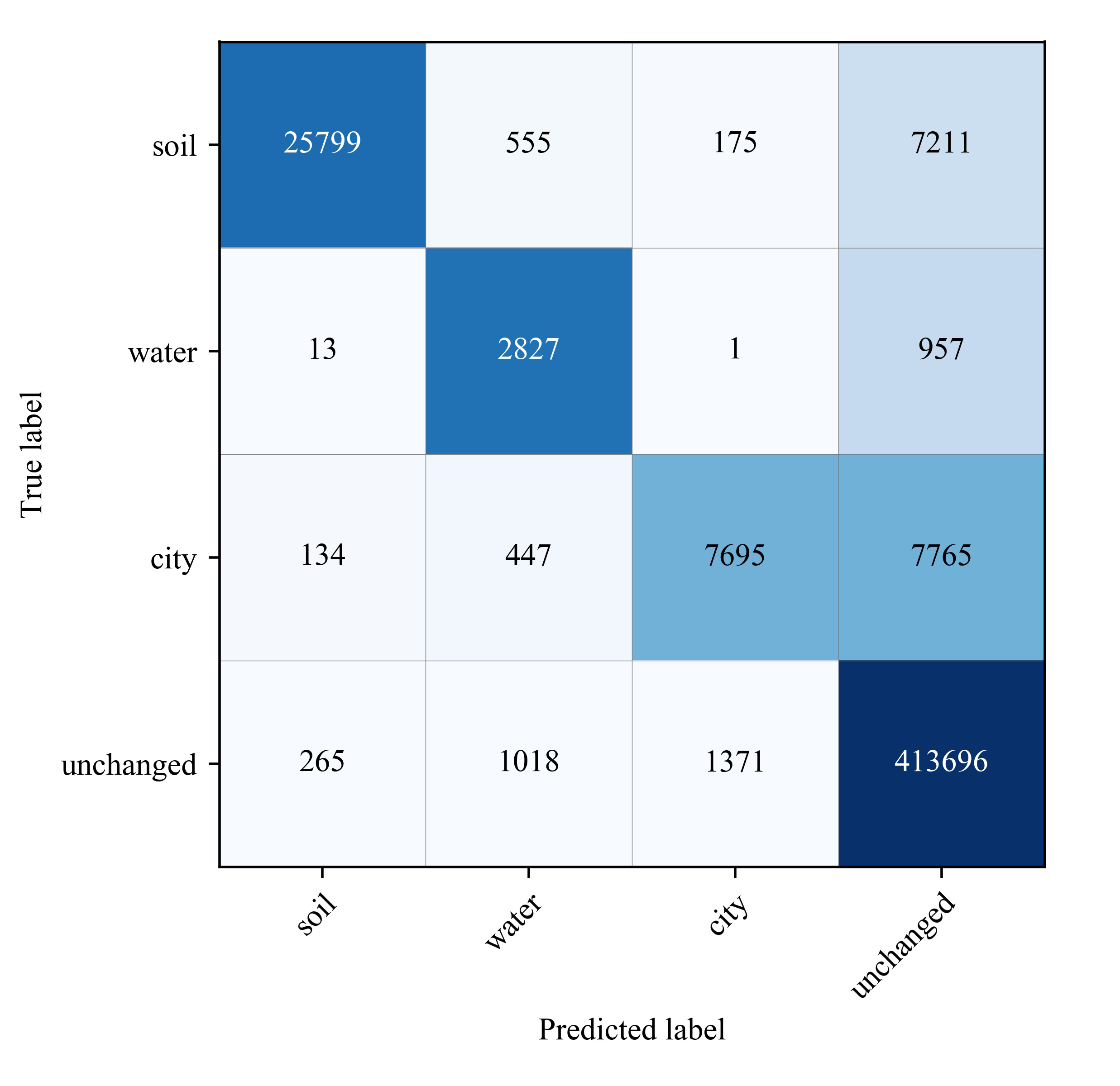}
  \label{fig_second_case}}
  \hfil
  \subfloat[]{
    \includegraphics[width=1.6in]{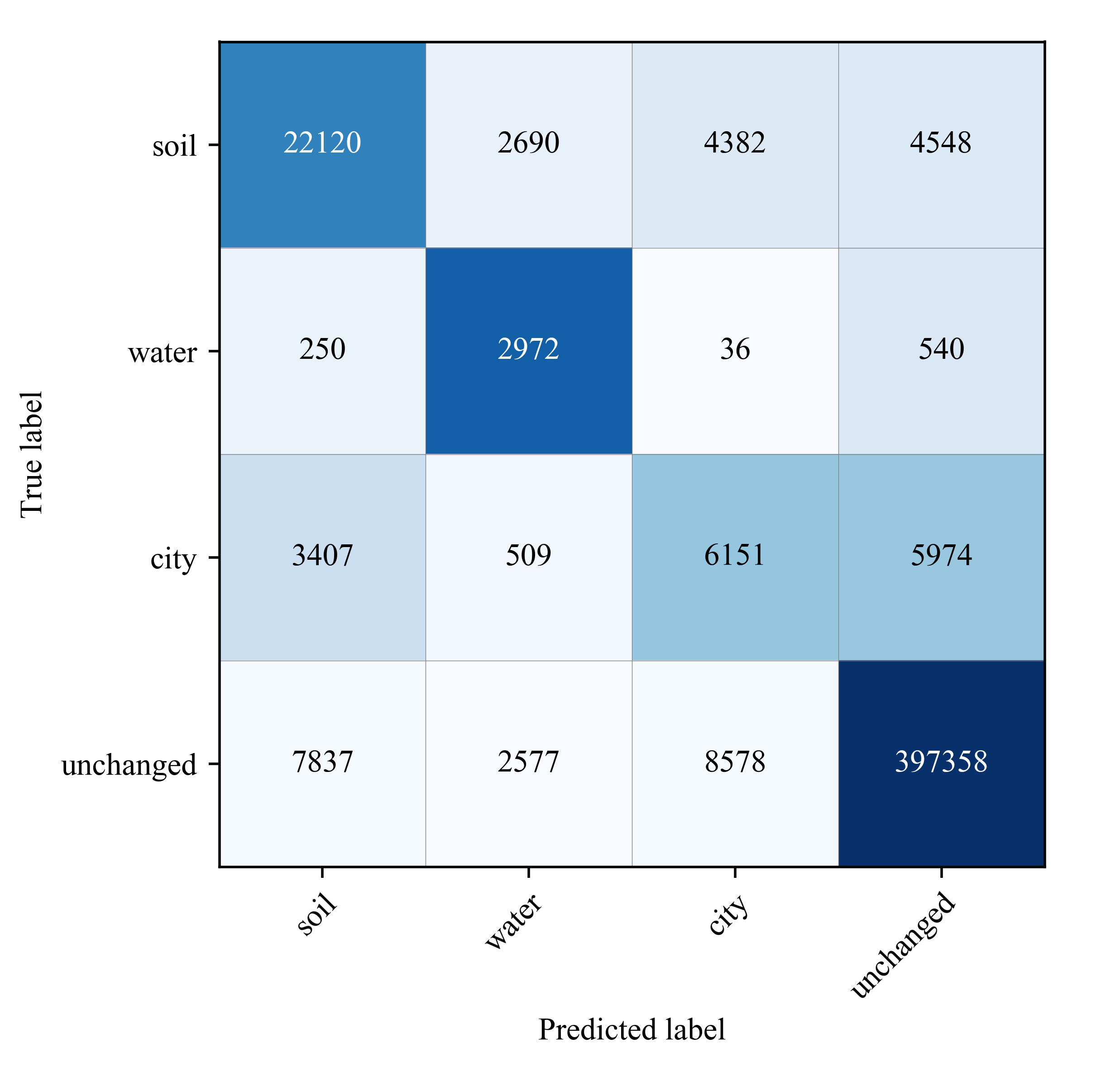}
  \label{fig_third_case}}
  \hfil
  \subfloat[]{
    \includegraphics[width=1.6in]{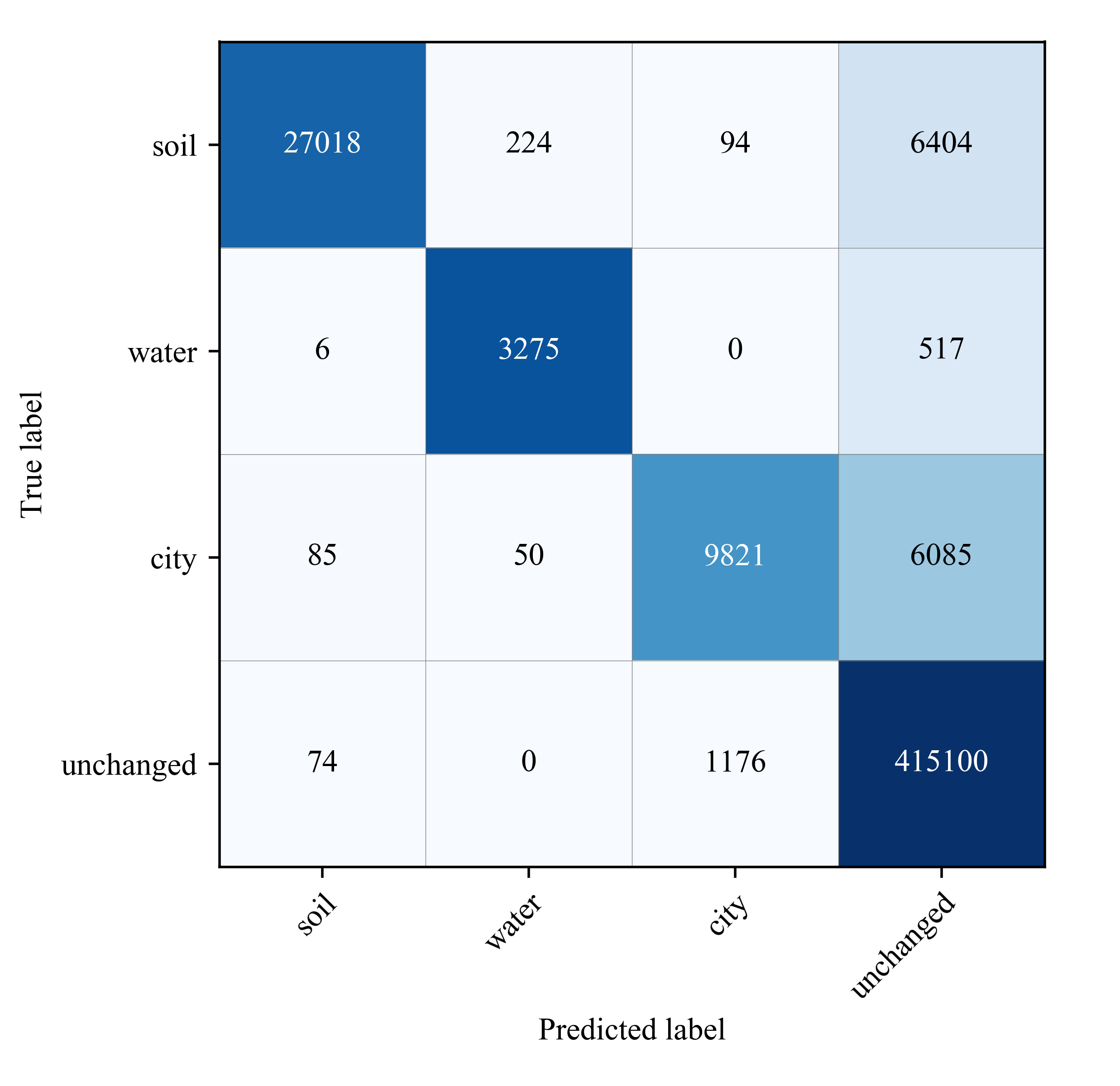}
  \label{fig_fourth_case}}
  
  \caption{Confusion matrices of the three comparison methods and KPCA-MNet. (a) $C^{2}$VA. (b) SAE. (c) RandNet. (d) KPCA-MNet.}
  \label{dis_confmat}
\end{figure*}

 The 2-D polar representations of change information extracted by $C^{2}$VA, SAE, RandNet, and KPCA-MNet are shown in Fig. \ref{dis_2dpolar}. And confusion matrices of the three comparison methods and KPCA-MNet are illustrated in Fig. \ref{dis_confmat}. As shown in Fig. 13-(a), in the polar domain of $C^{2}$VA, the three kinds of changes are separable from each other, but due to only original spectral information is used, some unchanged pixels are mixed with the pixels belonging to city changes and soil changes. For SAE, the spatial context information is utilized to get the spatial-spectral features, compared with $C^{2}$VA, the unchanged pixels are detected well, as shown in Fig. 14-(b). But the city changes are not highlighted in the magnitude, which leads to more than half of the pixels of city changes to be misclassified as unchanged pixels. As for RandNet, the unchanged pixels and changed ones are separable to a certain degree. However, since the random convolution kernels cannot extract discriminative features from VHR images, the three changes are mixed together and it is difficult to separate them from each other. Finally, in the 2-D polar domain of KPCA-MNet, it can be observed that the unchanged pixels are suppressed, changed pixels are highlighted and the three kinds of changes are compact and discriminative from each other in the WFD, which makes it easier to identify the changes and distinguish different change types. Compared with the three methods, each kind of changes and non-change are detected better by KPCA-MNet according to Fig. 14, which demonstrate that the spatial-spectral features extracted by KPCA-MNet are discriminative and more suitable for unsupervised multi-class CD. 

\par To verify the superiority of the WFD in multi-class CD, we compare it with other four representative methods. The first method, denoted as M-1, is performing clustering algorithms on the original feature difference map. The second method (M-2) is also performing clustering algorithms on the feature difference map, but the channels are weighted by the eigenvalues. The third way is similar to 2D-CVA. The two channels with the largest eigenvalues are selected from feature difference map and are adopted to calculate the direction, denoted as M-3. The last method entitled M-4 is adopting the direction measurement of $C^{2}$VA, which is similar to the proposed WFD, but the channels of feature difference map are not weighted by eigenvalues. The performance of these four methods and the proposed WFD is shown in Table \ref{mapping_table}. 

\begin{table}[t]
  \renewcommand{\arraystretch}{1.3}
  \caption{Performance comparison of different feature mapping methods in the HY data set}
  \label{mapping_table}
  \centering
  \begin{tabular}{c c c}
    \hline
    \bfseries   & \bfseries OA & \bfseries KC\\
    \hline\hline
    M-1	& 0.8136 &	0.4015 \\ 				
    M-2	&  0.9052	& 0.5867 \\ 												
    M-3	& \underline{0.9630}	& \underline{0.8028}	 \\ 												
    M-4 & 0.9447 &	0.7103 \\ 				
    Proposed	& \textbf{0.9687}	&	\textbf{0.8334} \\ 					 							
    \hline
  \end{tabular}
\end{table}

\par As we mentioned in section III-B, performing clustering algorithms in the original difference space is not effective, the KP of M-1 is only 0.4015. After weighted by eigenvalues, the KP of M-2 is increased by 46$\%$. However, the performance of M-2 is also unsatisfactory. By selecting the two channels containing the most change information to calculate the feature direction, the performance of M-3 is relatively good with OA of 0.9630 and KP of 0.8028. Though using all the channels, the performance of M-4 cannot compete with M-3. This is because the different channels of the feature difference map contain different amounts of change information, but M-4 treats these channels equally. By contrast, in the proposed WFD, the changed information in each channel is considered and the channel corresponding to the more changed information would play a more important role. Therefore, the proposed mapping way achieves the best performance, which proves the superiority of the WFD in multi-class CD. 

\section{Conclusion}\label{sec:6}
\par In this paper, a novel convolution operation based on kernel PCA, called KPCA convolution, is proposed for CD in multi-temporal VHR images. Compared with the conventional convolutional layer in CNN, the KPCA convolution can also extract nonlinear spatial-spectral features from VHR images, but the training process of the KPCA convolutional layer is an unsupervised learning fashion and does not require any annotated data. Based on the proposed KPCA convolution, a powerful and general network called KPCA-MNet is designed for unsupervised binary and multi-class CD in multi-temporal VHR images. In the KPCA-MNet, by the deep siamese network architecture consisting of several weight-shared KPCA convolutional layers, representative features are extracted from multi-temporal VHR images and two high-level spatial-spectral feature maps are generated. To train the network, an unsupervised layer-wise training way is developed. Through a pixel-wise subtraction operation, the feature difference map is generated. For the purpose of efficiently utilizing change information contained in the feature difference map, FM and WFD are proposed. The FM carries information about whether change exists and the WFD indicates what kind of change is. Then the feature difference map is mapped into a 2-D polar domain based on the FM and WFD. For the binary CD, a threshold segmentation method is adopted to get the result based on the FM. For the multi-class CD, a threshold segmentation method is first utilized to separate change and non-change, then a clustering algorithm is implemented on the WFD to get the multi-class change map. 

\par In the experiment of binary CD with the WH and QU data sets, the qualitative and quantitative results demonstrate that the proposed architecture outperforms the conventional CD models as well as deep learning-based methods. The experimental results of multi-class CD in the challenging HY data set confirm that the KPCA-MNet achieves better performance compared with several unsupervised multi-class CD methods. In addition, the proposed WFD can more efficiently utilize the multi-class change information in the feature difference map in contrast to other commonly used ways. Both experimental results reveal that the effectiveness of the proposed KPCA convolution and good versatility of KPCA-MNet in binary and multi-class CD. 

\par Our future work includes but is not limited to applying KPCA convolution and KPCA-MNet for heterogeneous image CD and hyperspectral image classification.

\ifCLASSOPTIONcaptionsoff
  \newpage
\fi

\bibliographystyle{IEEEtran}
\bibliography{KPCAMNet.bib}

\end{document}